\documentclass[prb,twocolumn,superscriptaddress,nobibnotes,amsmath,amssymb]{revtex4-1}
\usepackage{textcomp}
\usepackage{float}
\usepackage[utf8]{inputenc}
\usepackage[resetlabels]{multibib}
\usepackage{braket}
\usepackage{array}
\usepackage{graphicx}         
\usepackage{bm}               
\usepackage{color}
\definecolor{darkblue}{rgb}{0.0,0.0,0.3}
\definecolor{goodblue}{rgb}{0.0,0.0,0.6}
\usepackage{float}
\usepackage{hyperref}
\hypersetup{
	colorlinks   = true, 
	urlcolor     = darkblue, 
	linkcolor    = darkblue, 
	citecolor   = darkblue 
}

\makeatletter
\def\vhrulefill#1{\leavevmode\leaders\hrule\@height#1\hfill \kern\z@}
\makeatother

\newenvironment{stretchtext}
{\par\setlength{\parfillskip}{0pt}}
{\par}

\begin{document}
\thispagestyle{empty}
\onecolumngrid
\begin{center}
	\textbf{\large Direct Bandgap Emission from Hexagonal Ge and SiGe Alloys}
\end{center}
\begin{center}
\normalsize	{
			Elham~M.T.~Fadaly,$^{1,*}$ Alain~Dijkstra,$^{1,*}$ Jens~R.~Suckert,$^{2,*}$ Dorian~Ziss,$^{3}$ Marvin~A.J.v.~Tilburg,$^{1}$ Chenyang~Mao,$^{1}$ Yizhen~Ren,$^{1}$ Victor~T.v.~Lange,$^{1}$ Sebastian~K\"olling,$^{1,\ddagger}$ Marcel~A.~Verheijen,$^{1,5}$ David~Busse,$^{4}$ Claudia~R\"odl,$^{2}$ J\"urgen~Furthm\"uller,$^{2}$ Friedhem~Bechstedt,$^{2}$ Julian~Stangl,$^{3}$ Johnathan~J.~Finley,$^{4}$ Silvana~Botti,$^{2}$ Jos~E.M.~Haverkort,$^{1}$ Erik~P.A.M.~Bakkers$^{1,\dagger}$
			}
\smallskip
\small

\emph{$^\mathit{1}$Department of Applied Physics, Eindhoven University of Technology, Groene Loper 19, 5612AP Eindhoven, The Netherlands}
	
\emph{$^\mathit{2}$Institut f\"ur Festkörpertheorie und -optik, Friedrich-Schiller-Universität Jena, Max-Wien-Platz 1, 07743 Jena, Germany}
	
\emph{$^\mathit{3}$Institute of Semiconductor and Solid-State Physics, Johannes Kepler University, Altenbergerstra\ss~69, A-4040, Linz, Austria}
	
\emph{$^\mathit{4}$Physik Department, Walter Schottky Institut, Technische Universit\"at München, Am Coulombwall 4, 85748 Garching, Munich, Germany}

\emph{$^\mathit{5}$Eurofins Materials Science Netherlands BV, High Tech Campus 11, 5656 AE Eindhoven, The Netherlands}

\vspace*{0.5cm}
\end{center}

\twocolumngrid
\normalsize

\textbf{Silicon crystallized in the usual cubic (diamond) lattice structure has dominated the electronics industry for more than half a century. However, cubic silicon (Si), germanium (Ge) and SiGe-alloys are all indirect bandgap semiconductors that \textit{cannot} emit light efficiently. Accordingly, achieving efficient light emission from group-IV materials has been a holy grail \cite{Iyer1993} in silicon technology for decades and, despite tremendous efforts \cite{Miller1996,Ball2001,Canham2000,Green2001}, it has remained elusive \cite{Vivien2015}. Here, we demonstrate efficient light emission from direct bandgap hexagonal Ge and SiGe alloys. We measure a sub nanosecond, temperature insensitive radiative recombination lifetime and observe a similar emission yield to direct bandgap III-V semiconductors. Moreover, we demonstrate how by controlling the composition of the hexagonal SiGe alloy, the emission wavelength can be continuously tuned in a broad range, while preserving a direct bandgap. Our experimental findings are shown to be in excellent quantitative agreement with the \textit{ab initio} theory. Hexagonal SiGe embodies an ideal material system to fully unite electronic and optoelectronic functionalities on a single chip, opening the way towards novel device concepts and information processing technologies.}\par

Silicon has been the workhorse of the semiconductor industry since it has many highly advantageous physical, electronic and technological properties. However, due to its indirect bandgap, silicon \textit{cannot} emit light efficiently – a property that has seriously constrained potential for applications to electronics and passive optical circuitry \cite{Atabaki2018,Wang2018,Cheben2018}. Silicon technology can only reach its full application potential when heterogeneously supplemented \cite{Soref2016} with an efficient, direct bandgap light emitter.\par

\begin{figure*}[ht]
	\centering
	\includegraphics*[width=0.9\textwidth]{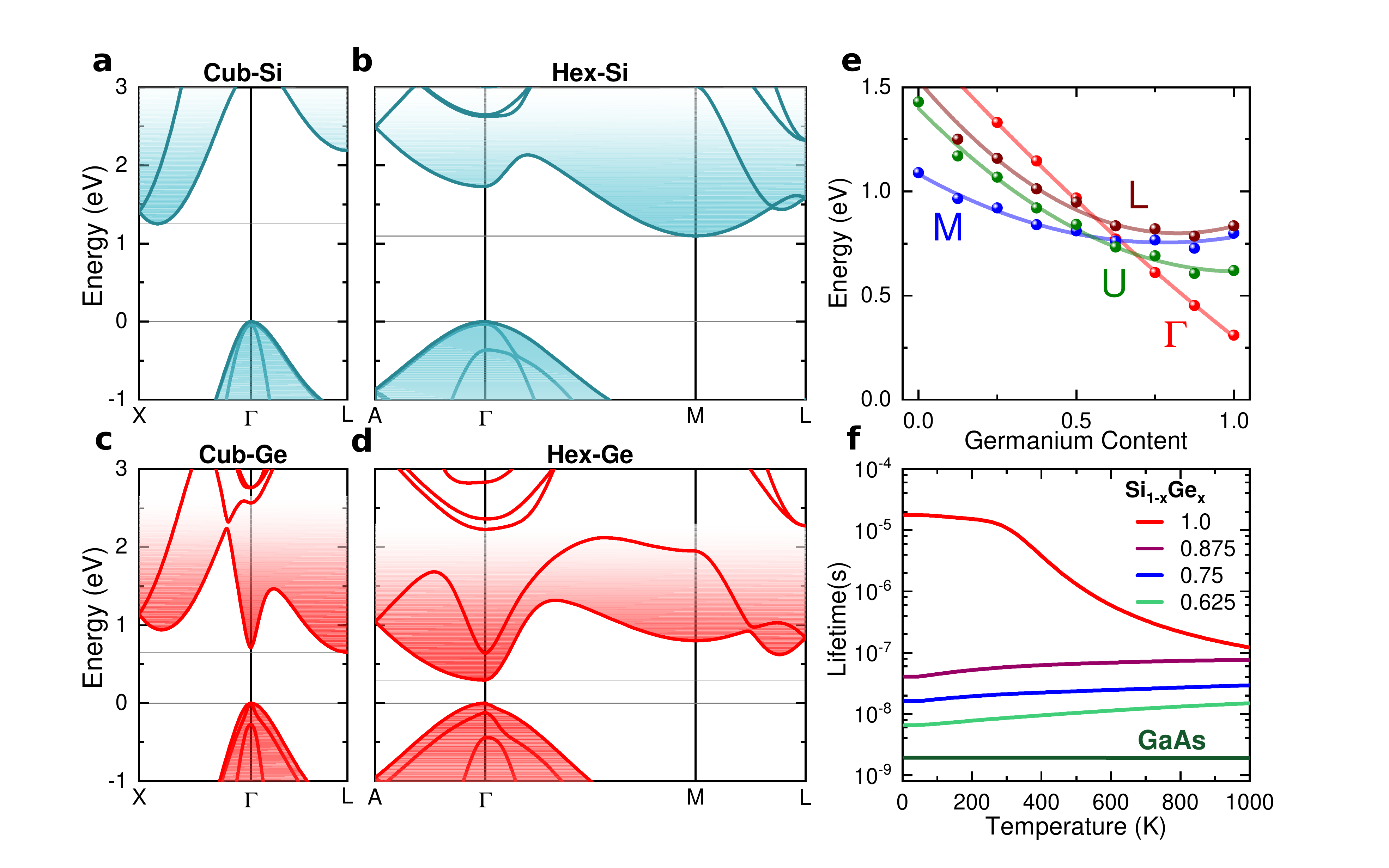}
	\caption{
		\textbf{Calculated band structure of Hex-Si$_{1-x}$Ge$_{x}$.} \textbf{a-d,} DFT calculations of the band structures of cubic  and Hex-Si and Ge. \textbf{e,} Energy of emission band minima fitted with a parabolic fit as a function of the Ge content in the Hex-Si$_{1-x}$Ge$_{x}$ alloy. \textbf{f,} Radiative lifetime of different Hex-Si$_{1-x}$Ge$_{x}$ compositions, with 10$^{19}$/cm$^{3}$ $n$-doping, as compared to the radiative lifetime of cubic GaAs.
	}\label{fig1}
\end{figure*}

The band structure of cubic Si, presented in Fig.~\ref{fig1}a is very well known, having the lowest conduction band (CB) minimum close to the X-point and a second lowest

\vspace{3.5pt}
\noindent \vhrulefill{0.25pt} \hspace*{7.2cm}\\

\footnotesize
\noindent $^*$ These authors contributed equally to this work.\\
$^{\dagger}$ Correspondence to E.P.A.M.(\href{mailto:e.p.a.m.bakkers@tue.nl}{e.p.a.m.bakkers@tue.nl}).\\
$^{\ddagger}$ Present address: Department of Engineering Physics, \'Ecole Polytechnique de Montr\'eal, C. P. 6079, Succ. Centre-Ville, Montr\'eal, Qu\'ebec H3C 3A7, Canada.

\normalsize
\noindent
minimum at the L-point.As such, it is the archetypal example of an indirect bandgap semiconductor, that, notwithstanding many great efforts \cite{Ball2001,Canham2000,Green2001,Vivien2015}, cannot be used for efficient light emission.By modifying the crystal structure from cubic to hexagonal, the symmetry along the $\langle111\rangle$ crystal direction changes fundamentally, with the consequence that the L-point bands are folded back onto the  $\Gamma$-point. As shown in Fig.~\ref{fig1}b, for hexagonal Si (Hex-Si) this results in a local CB minimum at the $\Gamma$-point, with an energy close to 1.7\,eV \cite{Raffy2002,De2014,Cartoixa2017}. Clearly, Hex-Si remains indirect since the lowest energy CB minimum is at the M-point, close to 1.1\,eV. Cubic Ge also has an indirect bandgap but, unlike Si, the lowest CB minimum is situated  at the L-point, as shown in Fig.~\ref{fig1}c. As a consequence, for Hex-Ge the band folding effect results in a direct bandgap at the $\Gamma$-point with a magnitude close to 0.3\,eV, as shown in the calculated band structure in Fig.~\ref{fig1}d \cite{Rodl2019}.\par

To investigate how the direct bandgap energy can be tuned by alloying Ge with Si, we calculated the band structures of Hex-Si$_{1-x}$Ge$_{x}$ (for $0<x<1$) using \textit{Ab initio} density functional theory (DFT) and a cluster expansion method for isostructural hexagonal binary alloys (see Methods). Selected results of our calculations, presented in Fig.~\ref{fig1}e, show the composition-dependent size of the emission bandgap for random Hex-Si$_{1-x}$Ge$_{x}$ alloys at high symmetry points in the Brillouin zone. Clearly, a direct bandgap is predicted at the $\Gamma$-point for $x>0.65$ (red curve) with a magnitude that is tunable across the energy range 0.3-0.7\,eV. This spectral interval is of technological interest for many potential applications including optical interconnects in computing \cite{Shen2019,Thomson2016}, silicon quantum photonic circuits \cite{Wang2018} and optical sensing \cite{Munoz2017,Wang2017}, among others \cite{Soref2015,Poulton2017}. Figure~\ref{fig1}f shows the calculated radiative lifetime of 10$^{19}$/cm$^3$ $n$-doped Hex-Si$_{1-x}$Ge$_{x}$ alloys for a polarization perpendicular to the c-axis, for different compositions. Remarkably, the radiative lifetimes of Hex-Si$_{1-x}$Ge$_{x}$ alloys are significantly lower than that of pure Hex-Ge, for which the lowest energy transition is dipole forbidden at the $\Gamma$-point \cite{Rodl2019}. This observation can be traced to the reduced symmetry in the random Hex-Si$_{1-x}$Ge$_{x}$ alloys, which leads to mixing of Ge s-states into the lowest conduction band wave function.We note, that the calculated lifetimes of the Hex-Si$_{1-x}$Ge$_{x}$ alloys are approaching those of III-V semiconductors, such as

\noindent
  GaAs (supplementary Table.~\ref{tab:st3}). Ge-rich alloys of Hex-Si$_{1-x}$Ge$_{x}$are thus highly appealing since they are theoretically predicted to combine a direct bandgap, strong optical transitions and wavelength tunability. Here, we demonstrate experimentally that Ge-rich alloys of Hex-Si$_{1-x}$Ge$_{x}$ are indeed direct gap semiconductors, observe strong emission, and a temperature independent nanosecond radiative lifetime. Our results are shown to be in remarkable quantitative agreement with theoretical predictions.\par

\begin{figure*}
	\centering
	\includegraphics[width=0.8\textwidth]{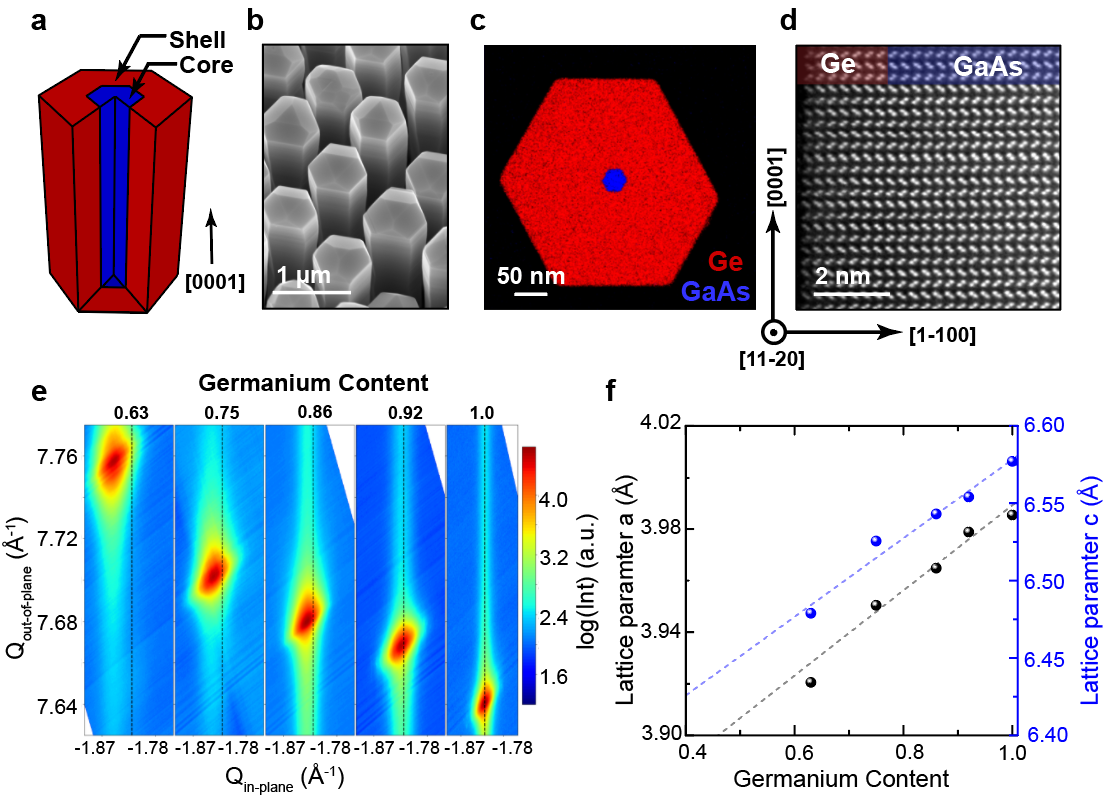}
	\caption{
	\textbf{Overview of the Hex-Si$_{1-x}$Ge$_{x}$ material system.} \textbf{a,} Schematic illustration of the hexagonal GaAs/Ge core/shell NWs drawn in blue/red, respectively. \textbf{b,} 30 \textdegree tilted view scanning electron micrograph of an array of epitaxial GaAs/Ge NWs grown on a GaAs (111)B substrate in the [0001] crystallographic direction. \textbf{c,} EDX image of a cross sectional lamella of a representative GaAs/Ge core/shell NW. \textbf{d,} Aberration-corrected HAADF-STEM image of the interface of GaAs/Ge structure obtained in the [11$\bar{2}$0] zone axis, displaying the ABAB stacking along [0001] of the hexagonal crystal structure. \textbf{e,} Reciprocal space maps around the hexagonal ($\bar{1}$018) NW reflections shown for five different Ge-concentrations. \textbf{f,} A plot of the Hex-Si$_{1-x}$Ge$_{x}$ in-plane ($a$) and out-of-plane ($c$) lattice parametersas a function of the Ge-content, the error is smaller than the data symbols (black and blue dots) used in the plot (see supplementary Table.~\ref{tab:st1}).
	}\label{fig2}
\end{figure*}

We begin by discussing the growth and crystalline properties of Hex-Si$_{1-x}$Ge$_{x}$ alloys. Various methods have been proposed to grow Hex-Si or -Ge including vapor-liquid-solid nanowire (NW) growth and strain-induced crystal transformation \cite{Vincent2014,Qiu2015,Pandolfi2018,He2019,Dushaq2019}. Recently, high quality Si-rich Hex-Si$_{1-x}$Ge$_{x}$ alloys have been grown using the crystal transfer method \cite{Hauge2015,Hauge2017} in which a wurtzite (WZ) gallium phosphide (GaP) core NW is used as a template for the growth of other materials as it is lattice matched to Si. Here, we grow Ge-rich Si$_{1-x}$Ge$_{x}$ alloys around a thin ($\sim 35$\,nm diameter) WZ gallium arsenide (GaAs) core that is lattice matched to Ge as shown in Fig.~\ref{fig2}a. We use a thin GaAs core to further reduce lattice strain and strain induced defects. The Au catalytic particles used to grow the WZ GaAs NW template have been removed by wet chemical etching and thick (200-400\,nm) Ge shells have been grown epitaxially on the WZ-GaAs (see Methods). The overview scanning electron microscopy (SEM) image presented in Fig.~\ref{fig2}b demonstrates that arrays of Hex-GaAs/Ge-core/shell NWs are formed on the growth substrate. These NWs are uniform in length and diameter and have smooth, well-defined {1100} side facets indicative of a single crystalline nature.Figure~\ref{fig2}c shows a cross-sectional Electron Dispersive X-ray (EDX) spectroscopy map confirming the expected core/shell geometry. The high-resolution High Angular Annular Dark Field (HAADF) Transmission Electron Microscopy (TEM) image presented in Fig.~\ref{fig2}d confirms the high-quality epitaxial growth of the Ge shell on the GaAs core and reveals an ABAB stacking along [0001]; the hallmark of a hexagonal crystal structure. These observations unequivocally confirm the single crystal nature of the NWs and their hexagonal crystal structure.\par
\stretchtext
The crystal quality and lattice parameters of a range of samples with GaAs/Si$_{1-x}$Ge$_{x}$ (with $x>0.5$) core/shell wires were studied by X-ray diffraction (XRD) measurements using synchrotron radiation. Fig.~\ref{fig2}e shows a set of asymmetrical reciprocal space maps (RSMs) for samples with Si$_{1-x}$Ge$_{x}$ shells with nominal Si compositions $x=1$, 0.92, 0.86, 0.75 and 0.63, respectively. The RSMs show the shift of the (1$\bar{0}$18) reflection that is exclusively allowed in the hexagonal crystal phase, as a function of the Ge-concentration. The higher the Ge-concentration, the more the hexagonal reflection shifts to lower 

\noindent
Q$_\mathrm{out-of-plane}$ and Q$_\mathrm{in-plane}$ shifts to lower Q$_\mathrm{out-of-plane}$ and Q$_\mathrm{in-plane}$ values, indicating an increase in the out-of-plane ($c$) and the in-plane ($a$) lattice parameters. From the narrow peak-width along Q$_\mathrm{out-of-plane}$, we can conclude that the overall crystal quality is very high, with an estimated stacking-fault (SF) density of 2-4\,SFs/$\mu$m along the crystalline [0001] direction. These results are in good agreement with the TEM measurements performed for the same samples (See supplementary Fig.~\ref{figs1}. We determine the $a$- and $c$-lattice parameters from a set of symmetric and asymmetric RSMs as a function of the Ge composition (See supplementary Fig.~\ref{figs1} and Table.~\ref{tab:st1}. The results of these experiments are presented in Fig.~\ref{figs2}f. Data points with $x>0.7$ lie on the linear interpolation line between Hex-Si and Ge (following Vegard’s rule) indicating that the lattice strain in the Si$_{1-x}$Ge$_{x}$ shell is negligible.\par

\begin{figure*}
	\centering
	\includegraphics[width=0.8\textwidth]{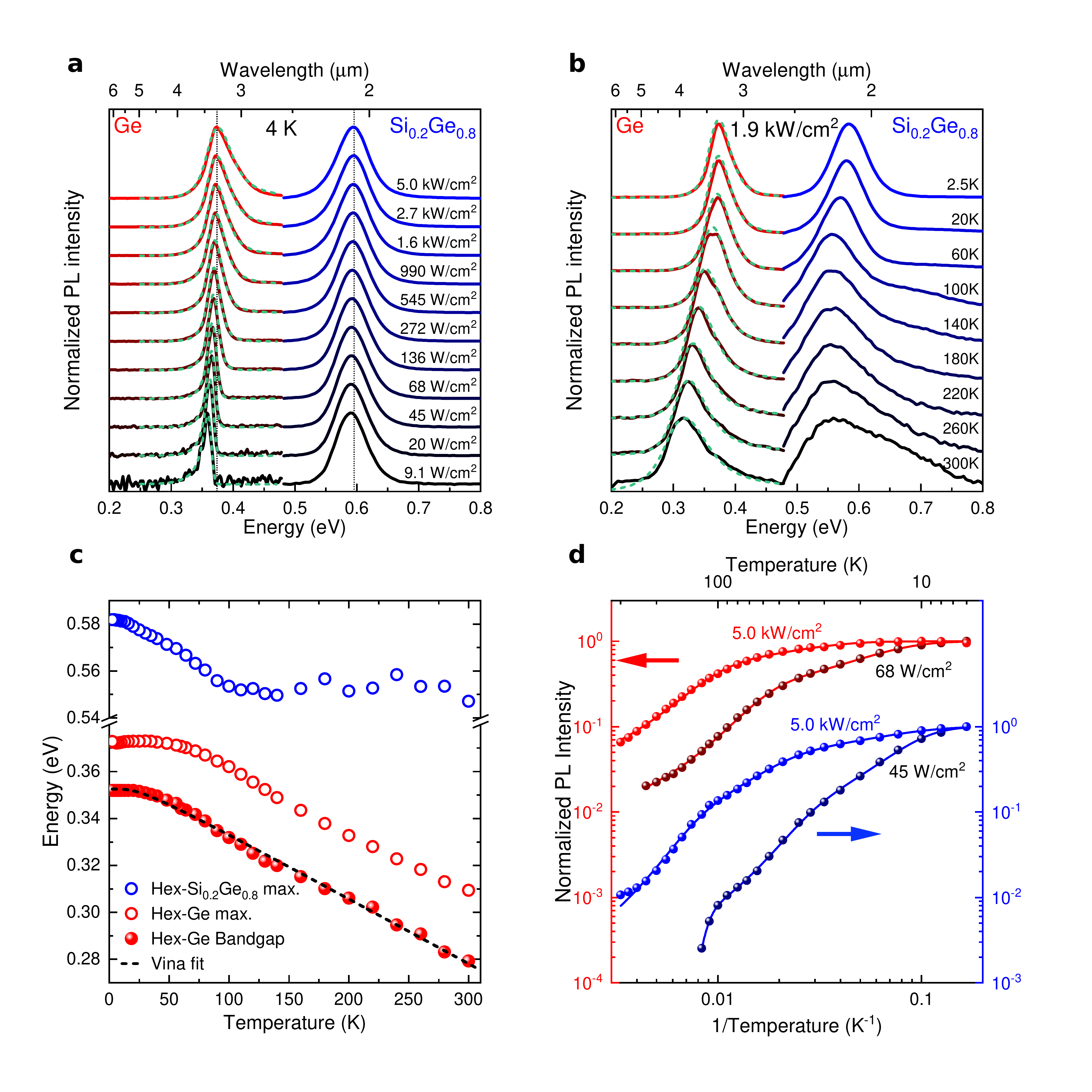}
	\caption{
		\textbf{Emission from Hex-Ge and Hex-Si$_{0.20}$Ge$_{0.80}$.} \textbf{a,} Excitation density dependent photoluminescence (PL) spectra of Hex-Ge (red to black) and Hex-Si$_{0.20}$Ge$_{0.80}$ (blue to black) as-grown samples, measured at 4\,K. All spectra are normalized to their own maximum. The LSW-fits of the Ge spectra are included as dashed lines. Vertical dotted black lines highlight the shift and broadening of the peaks, indicating BtB emission.\textbf{b,} Temperature dependence of the PL spectra, normalized to their own maximum, measured at an excitation density of 1.9\,kW/cm$^2$. A clear red shift and broadening are observed with increasing temperature, both indicating BtB recombination. The fits with the LSW model are shown as dashed lines. \textbf{c,} Shrinkage of the bandgap with temperature, fitted using the Vina equation (supplementary information, Section~\ref{section3}). The open circles show the maxima of the PL as plotted in (B) while the closed circles represent the bandgap determined by fits using the LSW model (supplementary information, Section~\ref{section2}). \textbf{d,} Arrhenius representation (supplementary information, Section~\ref{section4}) of the PL intensity as function of inverse temperature for Hex-Ge (red) and Hex-Si$_{0.20}$Ge$_{0.80}$ (blue). All intensities are normalized to their respective intensity at 4\,K. The reduced temperature dependence at higher excitation densities shows the approach towards the radiative limit.
	}\label{fig3}
\end{figure*}

We continue to explore the optical properties of the Hex-Si$_{1-x}$Ge$_{x}$ NWs probed using power and temperature dependent photoluminescence (PL) spectroscopy, shown in Fig.~\ref{fig3}a and ~\ref{fig3}b. We focus on two samples - pure Hex-Ge as the binary endpoint of the Hex-Si$_{1-x}$Ge$_{x}$ alloy and Si$_{0.20}$Ge$_{0.80}$  being representative of the binary alloy in the middle of the compositional range for which a direct bandgap is expected. Figure~\ref{fig3} presents power dependent PL spectra recorded at a temperature of 4\,K. The spectrum obtained from the Hex-Ge sample exhibits a narrow emission peak at the lowest excitation levels investigated. As the excitation density is increased, the emission peak broadens towards high energies and the peak blue-shifts by 19\,meV.\par

In order to understand the recombination mechanism, we have fitted both the excitation and temperature dependent data with the Lasher-Stern-Würfel (LSW) model \cite{P.Wurfel1982,Lasher1964} which describes band-to-band (BtB) recombination in a semiconductor. Model fits are included in Fig.~\ref{fig3}a and b, and confirm that the observed spectra of Hex-Ge can be explained by a BtB recombination process. From the fits, it can be concluded that the high energy broadening is due to an increase in the electron temperature and the observed blue-shift is due to the Burstein-Moss effect \cite{Moss1954}. In comparison to the pure Hex-Ge sample, the line width of the Hex-Si$_{0.20}$Ge$_{0.80}$ sample is larger due to alloy broadening (60\,meV compared to 14\,meV for Hex-Ge, at lowest excitation density) and can therefore not be fitted by the LSW model. Only a slight excitation-induced blue shift of 6\,meV was observed for the Si$_{0.20}$Ge$_{0.80}$ sample. Figure~\ref{fig3}b shows temperature dependent PL spectra recorded from the Hex-Ge and Si$_{0.20}$Ge$_{0.80}$ sample. Clear asymmetric broadening is observed at high temperatures, which, from the LSW model fits, can be assigned to broadening of the Fermi-Dirac distribution tail, supporting our identification that the observed emission peak is due to a BtB recombination pro-
\begin{figure*}
	\centering
	\includegraphics[width=0.8\textwidth]{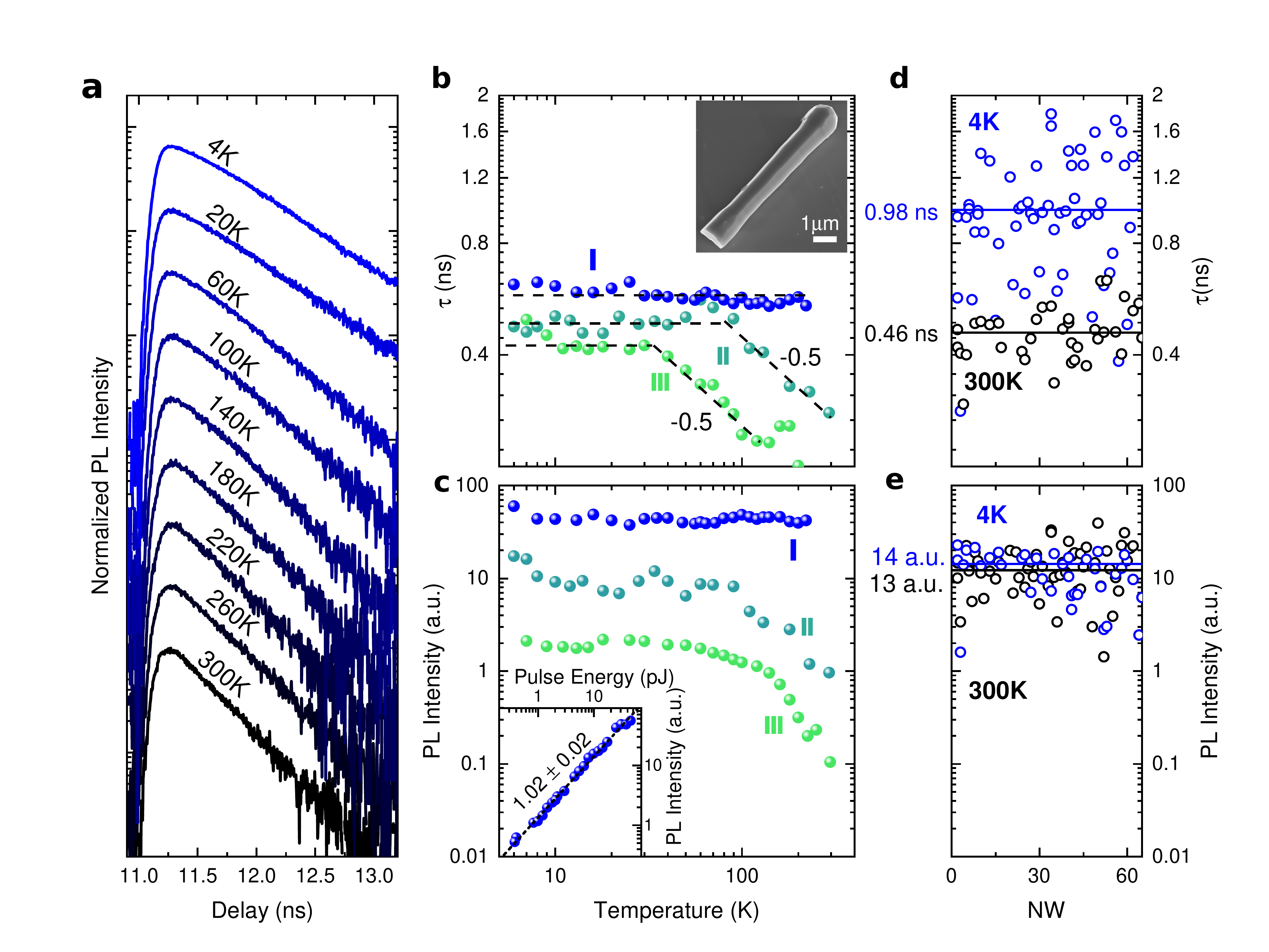}
	\caption{
        \textbf{Time-resolved PL measurements of single Hex-Si$_{0.20}$Ge$_{0.80}$ nanowires.}) \textbf{a,} PL lifetime measurements of Hex-Si$_{0.20}$Ge$_{0.80}$ recorded from a single wire for different temperatures from Sample I (see details in supplementary Table.~\ref{tab:st2}). All decay traces show a single exponential decay and are vertically shifted for clarity. \textbf{b,} Temperature dependence of the lifetime for three Hex-Si$_{0.20}$Ge$_{0.80}$ wires representing three different samples (I, II, III) with decreasing quality represented by blue to green colors. The onset of the reduction in lifetime due to non-radiative recombination shift to higher temperature for higher quality wires, as emphasized by the dashed lines. The inset shows a representative SEM image of a single NW from Sample I used for lifetime measurements. \textbf{c,} Integrated PL intensity as a function of temperature for the same wires as in panel (B), showing a nearly temperature independent radiative efficiency for the best sample (I, blue). The inset shows the excitation power dependence of the integrated PL intensity, exhibiting a slope very close to unity. \textbf{d,} Comparison of the low temperature (blue) and room temperature (black) lifetime for a set of ~60 wires from Sample I. The average lifetime shows a small decrease from 0.98\,ns at 4\,K to 0.46\,ns at 300\,K. \textbf{e,} Comparison of the integrated PL intensity at 4\,K and 300\,K of the same wires measured in (D), again showing a nearly temperature independent radiative efficiency.
	}\label{fig4}
\end{figure*}

 \noindent
 cess. The bandgap of Hex-Ge shifts from 3.5\,$\mu$m (0.353\,eV) at low temperature towards 4.4\,$\mu$m(0.28\,eV) at room temperature, confirming the expected bandgap shrinkage for a BtB transition as depicted in Fig.~\ref{fig3}c. The shrinkage of the Si$_{0.20}$Ge$_{0.80}$ bandgap as well as a detailed fit to the data of Hex-Ge, which yield a Debye temperature of 66\,K, is discussed in (supplementary information, Section~\ref{section3}). Figure~\ref{fig3}d shows the temperature dependence of the integrated emission intensity of the samples on an Arrhenius representation. A decrease (factor 15-100) of the integrated emission intensity is observed upon increasing the lattice temperature. The ratio of the photoluminescence emission intensity at 4\,K and 300\,K compares favorably to many well-developed III-V semiconductors (see supplementary information, Section~\ref{section7}). The decrease of the intensity with increasing temperature is suppressed for higher excitation powers, as shown in Fig.~\ref{fig3}d, due to saturation of non-radiative processes. The fact that the emission decreases with increasing temperature provides the first indication that Hex-Ge is a direct band gap semiconductor. In contrast, for an indirect gap semiconductor at low temperature, excited carriers accumulate in the indirect minimum and do not, therefore, efficiently emit light. As the lattice temperature increases, the photoluminescence intensity is expected to increase \cite{Wirths2015} as carriers are thermally excited into the higher energy direct minimum from where they can recombine with a higher quantum efficiency.\par
\begin{figure*}
\centering
\includegraphics[width=0.9\textwidth]{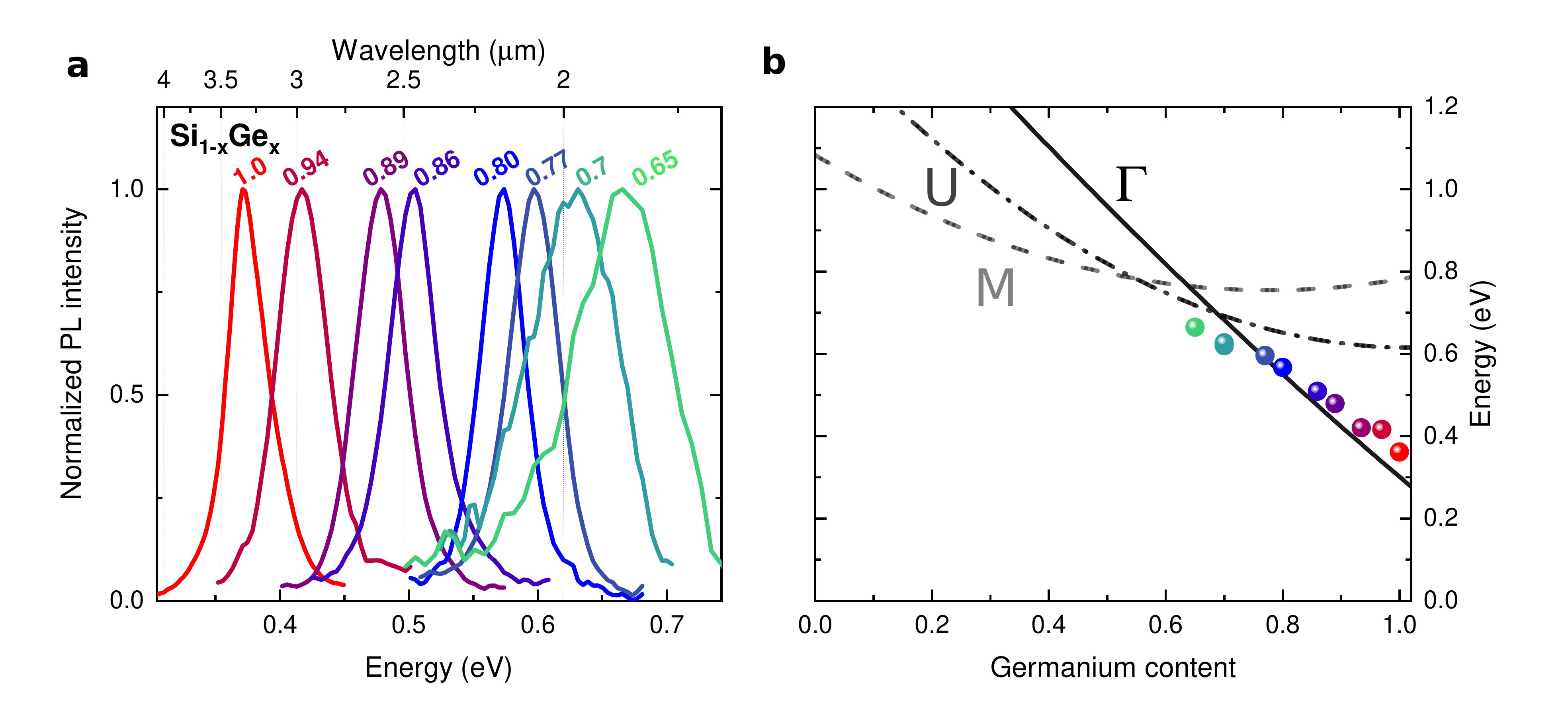}
\caption{
	\textbf{Tunability of the direct bandgap of Hex-Si$_{1-x}$Ge$_{x}$ alloy.} \textbf{a,} Tunability of the PL spectra for different compositions. The spectra were recorded at 4\,K at an excitation density of 1.9\,kW/cm$^{2}$ on as-grown samples. The spectra for Hex-Si$_{0.30}$Ge$_{0.70}$ and Hex-Si$_{0.35}$Ge$_{0.65}$ where measured  on wires dispersed onto a silicon substrate capped with a gold (Au) layer, see Methods. \textbf{b,} Comparison of the measured peak energy as a function of the Ge content with the calculated emission band minima.
}\label{fig5}
\end{figure*}

\indent
We next deduce the radiative lifetime as well as the radiative emission efficiency of Hex-Si$_{0.20}$Ge$_{0.80}$. It is important to note that the measured decay lifetime is determined by the fastest recombination process, which can be either radiative or non-radiative in nature. It is therefore crucial to choose experimental conditions in which the measured recombination lifetime is exclusively governed by pure radiative recombination. This can be achieved at low temperature, since non-radiative processes are commonly thermally activated and therefore negligible, see supplementary information, Section~\ref{section6}. Moreover, we choose to measure at high excitation density in order to saturate nonradiative processes and to maintain the radiative limit up to increased temperature. Typical results from time-resolved luminescence measurements on a single wire from the Si$_{0.20}$Ge$_{0.80}$ Sample are presented in Fig.~\ref{fig4}a as a function of temperature. We observe a clear mono-exponential decay transient, characteristic of a single, dominant decay channel. For all Hex-Si$_{1-x}$Ge$_{x}$ NWs investigated, the characteristic recombination lifetime is around 1\,ns, very similar to conventional direct gap semiconductors such as GaAs or InP at low temperatures with similar doping levels (supplementary information, Section~\ref{section7}). We observe that the experimentally obtained lifetime is an order of magnitude smaller than the theoretically calculated lifetime, which indicates that the perfect crystal symmetry is also broken by other factors \cite{Rodl2019}. Figures~\ref{fig4}b and c show the temperature dependence of both the recombination lifetime and the integrated emission intensity as a function of temperature from single wires from 3 different samples grown under different conditions leading to different quality, see supplementary Table.~\ref{tab:st2}. The wires show comparable lifetimes at low temperature, but the intensity and lifetime both start to decrease at a temperature of around 40\,K (100\,K) for sample III (II) which is the low (medium) quality wire. For the higher quality sample, we observe a constant integrated photoluminescence intensity and lifetime as a function of temperature up to 220\,K, which conclusively shows the absence of non-saturated thermally activated non-radiative recombination processes, providing strong evidence for pure radiative recombination, see Supplementary Information sections~\ref{section4} and 4 and \ref{section7}.\par

In order to be sure that the data for an individual wire are representative, we analyzed more than 60 individual wires swiped from the high crystal quality Sample I, which are presented in Fig.~\ref{fig4}e and d. The analysis shows that both the photoluminescence efficiency and the lifetime are almost temperature independent up to 300\,K. We subsequently analyze the excitation power dependence of the emitted photoluminescence intensity in Fig.~\ref{fig4}c inset. Importantly, the plot shows a linear increase of the photoluminescence intensity with a slope very close to unity, which is consistent with a pure radiative decay mechanism, see supplementary information, Section~\ref{section5}. Since our measurements are performed in the radiative limit and the carriers accumulate in the direct minimum at low temperature, we conclude that we observe direct band gap emission with a sub-nanosecond recombination lifetime.\par 

The combination of our theoretical predictions, structural microscopy data and luminescence data shows conclusive evidence for Hex-Si$_{1-x}$Ge$_{x}$ $(0.65<x<1)$ being a new class of direct bandgap semiconductors with a large optical matrix element. We subsequently compare the radiative transition rate of Hex-SiGe with other direct bandgap semiconductors. The radiative transition rate $R_{rad}$ is quantified by $R_{rad}=B_{rad}np$ in which $n$ and $p$ are the electron and hole densities and $B_{rad}$ is the coefficient for radiative recombination, which is directly related to the transition dipole moments. The coefficient $B_{rad}$ can be deduced from a measurement of the pure radiative lifetime, $\tau_{rad}$ by $B_{rad}=1/(\tau_{rad}n_{0})$ in which $n_0$ is the activated donor density. As explained in the supplementary information, Section~\ref{section8}, we obtain 0.7$\times$10$^{-10}$\,cm$^{3}$/$s<B_{rad}<$1$\times$10$^{-10}$\,cm$^{3}$/s at 300\,K, which is comparable in magnitude to GaAs \cite{Lush2009} and InP \cite{Semyonov2010} and almost 5 orders of magnitude larger than for cubic Si \cite{Trupke2003}, see supplementary Table.~\ref{tab:st3}. Hex-Si$_{1-x}$Ge$_{x}$ is thus a fully silicon compatible semiconductor with a radiative emission strength comparable to a direct bandgap III-V semiconductor.\par
 
Now that we have established the direct nature of the bandgap of Hex-Si$_{1-x}$Ge$_{x}$, we demonstrate how the size of the direct bandgap can be tuned via compositional engineering. Figure~\ref{fig5} shows PL measurements recorded at T=4\,K from the series of samples with $x=0.65-1.00$. Bright emission is observed that red shifts with increasing Ge-content from 0.67 eV ($x=0.65
$) to 0.35 eV ($x=1.00$). The peak energy of the emission is compared in Fig.~\ref{fig5}b with the calculated energy of the direct bandgap ($\Gamma$) revealing excellent agreement. Clearly, the measured transition energies agree remarkably well with our theoretical predictions. The excellent agreement between theory and experiment provides not only a verification of the calculated bandgaps, but also provides a strong support for the existence of a direct bandgap in Hex-Si$_{1-x}$Ge$_{x}$ for $x>0.65$.\par

 Direct bandgap Hex-Si$_{1-x}$Ge$_{x}$ opens a pathway towards tight monolithic integration \cite{Miller2017} of Hex- Si$_{1-x}$Ge$_{x}$ light sources with passive cubic Si-photonics circuitry \cite{Atabaki2018,Wang2018,Cheben2018} on the same chip. This will reduce stray capacitances thereby increasing performance and reducing energy consumption which is important for green information and communication technologies. Now that the fundamental boundaries have been removed, a challenge is the development of a Hex-Si$_{1-x}$Ge$_{x}$ technology platform on conventional Cubic Si substrates. Possible integration routes are strain-induced transformation \cite{Qiu2015} of Si$_{1-x}$Ge$_{x}$, for instance by a dielectric (i.e. SiO$_x$ or SiN$_x$) strain envelope, or alternatively by template-assisted selective area growth \cite{Staudinger2018}of the hexagonal phase.

\vspace*{2\baselineskip}
{\centering	\textbf{Methods}\par}
\vspace*{1\baselineskip}
\noindent \textbf{Ab initio Calculations.}
All calculations were performed using density functional theory (DFT) as implemented in the Vienna \textit{Ab initio} Simulation Package (\textsc{Vasp})\cite{Kresse1996} with the projector augmented wave method. We used a plane-wave cutoff of 50\,meV and we included Ge 3d electrons as valence electrons. Brillouin zone integrations were carried out using $12\times12\times6$ $\Gamma$-centered \textbf{k}-point grids for lonsdaleite Ge and $12\times12\times6$ $\Gamma$-centered $k$-point grids for Si-Ge, ensuring a convergence of total energies to 1\,meV/atom. For structural calculations, the PBEsol exchange-correlation potential\cite{Perdew2008} was used, together with a convergence threshold of 1\,meV/\AA\, on Hellmann-Feynman forces. The modified Becke-Johnson exchange potential in combination with local density approximation (MBJLDA)\cite{Tran2009} was preferred for electronic structures and optical properties, as it guarantees bandgaps in excellent agreement with experiments\cite{Borlido2019}. We included spin-orbit coupling in all calculations. More details on the \textit{Ab initio} method and the selected approximations can be found in Ref.\cite{Rodl2019}.\\
Alloys are studied  using a cluster expansion method for isostructural lonsdaleite binary alloys\cite{Schleife2011}. For the cluster expansion, the macroscopic alloy is divided into clusters of 8 atoms obtained from the primitive wurtzite (WZ) unit cell. In this way, it is possible to study 46 different structures ranging from pure Ge to pure Si. This method becomes more accurate with increasing size of the clusters, and we verified that the thermodynamic averages are not significantly modified by performing calculations with 16 atom clusters. The radiative lifetime ($\tau_{rad}$) at temperature ($T$) is calculated using the formula:\\

\begin{equation}\label{eq:1_tau}
\frac{1}{\tau_{rad}}= \sum\limits_{cv\textbf{k}} A_{cv\textbf{k}}\ w_\textbf{k} f_{c\textbf{k}} (1-f_{v\textbf{k}})
\end{equation}

where $A_{cv\textbf{k}}$ denotes the radiative recombination rate for vertical optical transitions between a conduction state $\ket{c\textbf{k}}$ and a valence state $\ket{v\textbf{k}}$, with one-particle energies $\epsilon_{c\textbf{k}}$ and $\epsilon_{v\textbf{k}}$, and Fermi occupation functions $f_{c\textbf{k}}$ and $f_{v\textbf{k}}$ and $\textbf{k}$-point weight $w_\textbf{k}$. In order to reproduce experimental conditions, we included $n=10^{19}\,cm^{-3}$ charge carriers due to $n$-doping in the conduction band, and modified accordingly the chemical potential of electrons. The radiative recombination rate is given by:\\
\begin{equation} \label{eq:Acvk}
    A_{cv\textbf{k}} = n_\mathrm{eff}\, \frac{e^2\, (\epsilon_{c\textbf{k}}-\epsilon_{v\textbf{k}})}{\pi \epsilon_0\,\hbar^2 m^2 c^3}\, \frac{1}{3}\sum\limits_{j=x, y, z} \left|\braket{c\textbf{k}|p_j|v\textbf{k}}\right|^2
\end{equation}

where $n_{eff}$ is the refractive index of the effective medium (here set approximately to the experimental value for cubic Ge for which $n_{eff}$=5) The squares of the momentum matrix elements can be either averaged over all directions corresponding to the emission of unpolarized light, as in Eq.~\eqref{eq:Acvk}, or only the in-plane component is considered, for light polarized perpendicularly to the wire axis. Denser k-point grids were necessary to calculate lifetimes ($72\times72\times36$ for lonsdaleite Ge and $24\times12\times12$ for Si-Ge).\\

\noindent \textbf{Materials Synthesis.}
The GaAs NWs were grown in a Close Coupled Shower head (CCS) Metal Organic Vapor Phase epitaxy (MOVPE) reactor via catalyst assisted growth following the Vapor-Liquid-Solid (VLS) mechanism utilizing gold (Au) catalyst seeds as schematically illustrated in supplementary Fig.~\ref{figs1}. The Au catalyst seeds were deposited in nano disks arrays arrangement on a GaAs (111)B substrate via the electron beam lithography technique. The growth was performed at a reactor flow of 8.2\,slm (standard litres per minute) and a reactor pressure of 50\,mbar. For the GaAs NWs, the growth template was annealed at a set thermocouple temperature of 635\,\textdegree C under an AsH$_3$flow set to a molar fraction of $\chi_\mathrm{AsH_3}=6.1\times10^{-3}$. Then, the growth was performed at a set temperature of 650℃ with trimethylgallium (TMGa) and Arsine (AsH$_3$) as material precursors set to molar fractions of $\chi_\mathrm{TMGa}=1.9\times10^{-5}$, $\chi_\mathrm{AsH_3}=4.55\times10^{-5}$, respectively, resulting in  a V/III ratio of 2.4. After the growth of the GaAs core NWs, they are chemically treated with a cyanide based solution to remove the Au catalyst particles to avoid gold contamination in the SiGe shells, (see, supplementary Fig.~\ref{fig3}c). Eventually, the GaAs NW core is used as a hexagonal material template and was overgrown with a  Si$_{1-x}$Ge${x}$ shell by introducing the suitable gas precursors for the shell growth which are GeH$_4$ and Si$_2$H$_6$. The Si$_{1-x}$Ge$_{x}$ shell was grown at a set temperature of 650-700\,\textdegree C at a molar fraction of $\chi_{SiGe}=1.55\times10^{-4}$ for a certain time according to the desired thickness.\\

\noindent \textbf{Structural Characterization.}
The structural quality of the crystals was investigated by transmission electron microscopy (TEM). Two different sample preparation methods were used. In the standard axial analysis, NWs were mechanically transferred to a holey carbon TEM grid. Concerning the cross-section TEM studies, NWs were prepared using a Focused Ion Beam (FIB). In both cases, high resolution TEM and Scanning TEM analyses were conducted using a JEM ARM200F probe-corrected TEM operated at 200\,kV. For the chemical analysis, Electron Dispersive Xray (EDX) spectroscopy measurements were carried out using the same microscope equipped with a 100\,mm$^2$ EDX silicon drift detector. TEM lamellae were prepared in a FEI Nova Nanolab 600i Dual beam system. For this, the NWs were initially transferred with the aid of a Kleindiek nano-manipulator from the growth substrate to a piece of Si and then arranged to lie parallel to each other. These NWs were covered with electron- and ion-beam induced metal deposition to protect them during the procedure. The lamella was cut out by milling with 30\,kV Ga ions and thinned down with subsequent steps of 30, 16, and 5\,kV ion milling in order to minimize the Ga-induced damage in the regions imaged with TEM.\\

\noindent \textbf{Atom Probe Tomography.}
For the APT measurements, individual nanowires (NWs) were isolated from a forest of NWs as described previously\cite{koelling2017atom} with a Kleindiek nano-manipulator inside a FEI Nova Nanolab 600i Dual beam. APT analyses were carried out in a LEAP 4000X-HR from Cameca. The system is equipped with a laser generating picosecond pulses at a wavelength of 355\,nm. The experimental data were collected at laser or voltage pulse rates between 65-125 kHz with laser pulse energies between 5-10\,pJ or pulse fractions between 25-27.5\%. No significant differences between laser and voltage pulses are seen aside from a slightly higher compression of the core in laser pulsed mode as discussed in Ref8 and a lower quality of the mass spectra in voltage pulsed mode. During the analysis the sample is kept at a base temperature of 20\,K in a vacuum of ~$2\times10^{-11}$\,mbar. Details of the APT measurement are explained in Ref.\cite{koelling2016impurity}.\\

\noindent \textbf{X-Ray Diffraction.}
The XRD measurements have been carried out at the Deutsches -Elektronen - Synchrotron (DESY) in Hamburg, at the high-resolution-diffraction beamline P08. For the diffraction experiments a high precision 6-circle diffractometer has been used, the photon energy was set to 15\,keV with a corresponding wavelength of 0.8266\,\AA. The energy was carefully chosen to ensure a high photon flux while still being able to access higher-indexed reflections, needed for the precise measurements of the lattice parameters. The x-ray beam has been shaped by a slit system and the resulting spot-size on the sample was 200\,$\mu$m (horizontal)$\times$100$\mu$m (vertical), a size sufficient to illuminate a few thousands of wires at once. For measuring the scattered signal coming from the wires, a Dectris – “Mythen” 1D X-ray detector has been used; this detector offers a high dynamic range and, due to the small pixel size (50\,$\mu$m), an increased angular resolution in 2$\theta$, compared to most 2D detectors. For the conversion of the measured angular coordinates to reciprocal space coordinates and all further data processing, such as 2D-peak-fitting and post-processing for plotting, the freely available software library “Xrayutilities” in combination with Python 3.6 has been used\cite{Kriegner:rg5038}.\\

\noindent \textbf{Optical Characterization.}
Time-correlated single photon counting measurements have been performed on single Si$_{0.20}$Ge${0.80}$ wires. The wires have been mechanically transferred onto a silicon wafer with a chromium (15\,nm), Gold (300\,nm) and SiO$_x$ (12\,nm) top layer to act as a back mirror. This approach enhances the measured intensity and avoids potential optical signals emitted by the wafer. 
The samples with transferred Si$_{0.20}$Ge$_{0.80}$ wires were mounted in an Oxford Instruments HiRes2 helium flow cryostat and were excited with a 1030\,nm, NKT ONEFIVE Origami femto-second pulsed laser with a 40\,MHz repetition rate. The photoluminescence signal was measured in a backscattering geometry using a 36X gold coated cassegrain objective which focused the excitation laser to a spot of $\approx$\,3\,$\mu$m. The laser was filtered out of the PL signal using a 1350\,nm long pass filter. Using an achromatic lens the PL signal was then focused onto a SM2000 single mode fiber and fed to a Single Quantum superconducting-nanowire-single-photon-detector which was optimized for $a>35\%$ quantum efficiency at 1800\,nm and $a>15\%$ quantum efficiency at 2000\,nm. The 1350\,nm long pass filter in combination with the SM2000 fiber defined a spectral interval of 1350\,nm to $\approx$\,2300\,nm over which PL was integrated. The time correlations between a laser pulse and a detection event were measured and counted using a PicoQuant PicoHarp 300 module. The overall instrumental response function (IRF) shows a FWHM of 96\,ps with a decay time of $\tau_{IRF}$=21\,ps which is the minimum observable decay time of the system. All measurements presented in Fig.~\ref{fig4} have been performed with 125\,pJ pulses resulting in an excitation density of $\approx$\,0.4\,mJ/cm$^2$/pulse, with the exception of the inset of Fig.~\ref{fig4}c where the excitation energy was varied. All lifetime measurements have been baseline corrected and fitted using a single exponential decay transient.\\
Spectrally resolved Photoluminescence experiments with accurate temperature control have been carried out on as-grown samples mounted in an Oxford Instruments HiRes2 helium flow cryostat. The samples were illuminated using a 976\,nm, continuous wave laser, modulated at a frequency of 35\,kHz, focused down to a 45\,$\mu$m spot on the sample using a 2.1\,cm focal distance off-axis parabolic Au-mirror. The same off-axis parabolic mirror was used to collimate the photoluminescence signal and coupled it into a Thermo Scientific Nicolet IS50r FTIR, equipped with an MCT detector, used for Si$_{1-x}$Ge$_{x}$ samples with $x>0.80$ and an extended-InGaAs detector, used for samples with $x\leq0.80$. The FTIR was operated in step-scan mode, which allowed to use a lock-in technique to eliminate the thermal background. In order to minimize parasitic absorption, the full optical path was purged with nitrogen.

\bibliography{MainText}

\vspace*{2\baselineskip}
{\centering	\textbf{Acknowledgments}\par}
\vspace*{1\baselineskip}
We thank Dan Buca, Antonio Polimeni , Jo de Boeck and Chris Palmstrøm for their valuable feedback on the manuscript. Also, we thank Rene van Veldhoven for the technical support of the MOVPE reactor. Furthermore, we acknowledge DESY (Hamburg, Germany), a member of the Helmholtz Association HGF, for the provision of experimental facilities. Parts of this research were carried out at PETRA III and we would like to thank Florian Bertram for assistance in using beamline P08. Funding: This project has received funding from the European Union’s Horizon 2020 research and innovation program under grant agreement No 735008 (SiLAS), the Dutch Organization for Scientific Research (NWO) and from Marie Sklodowska Curie Action fellowship (GA No. 751823); We acknowledge Solliance, a solar energy R\&D initiative of ECN, TNO, Holst, TU/e, imec, Forschungszentrum Jülich, and the Dutch province of Noord-Brabant for funding the TEM facility. We thank the Leibniz Supercomputing Centre for providing computational resources on SuperMUC (Project No. pr62ja).

\vspace*{2\baselineskip}
{\centering	\textbf{Author contributions}\par}
\vspace*{1\baselineskip}

 E.F., C.M. and Y.R. carried out the growth of WZ NW cores. E.F. carried out the growth of SiGe shells and analyzed the data. A.D. and D.B. carried out the photoluminescence spectroscopy. A.D. analyzed the optical data. M.v.T., A.D., and V.v.L performed time-resolved spectroscopy on single NWs; J.S., C.R., J.F. and S.B. performed the DFT calculations D.Z. and J.S. performed the XRD measurements; S.K. performed the APT characterization; M.V. performed the TEM analysis; J.S., J.F.F., S.B., J.H., and E.B. supervised the project; F.B. contributed to the interpretation of data and E.F., A.D., D.Z., S.B., J.F.F., J.H. and E.B. contributed to the writing of the manuscript. All authors discussed the results and commented on the manuscript.

\thispagestyle{empty}
\onecolumngrid
\hypersetup{urlcolor=darkblue}
\clearpage
\onecolumngrid
\setlength\parindent{0pt}
\linespread{1.2}
\setcounter{equation}{0}
\renewcommand{\theequation}{S\arabic{equation}}
\renewcommand{\eqref}[1]{equation\ref{#1}} 
\renewcommand{\thefigure}{S\arabic{figure}}
\renewcommand{\thetable}{S\arabic{table}}
\renewcommand{\thesection}{S\arabic{section}}
\setcounter{figure}{0}

\begin{center}
	\textbf{\large Supplementary Information: Direct Bandgap Emission from Hexagonal Ge and SiGe Alloys}
\end{center}
\begin{center}
	\normalsize	{
			Elham~M.T.~Fadaly,$^{1,*}$ Alain~Dijkstra,$^{1,*}$ Jens~R.~Suckert,$^{2,*}$ Dorian~Ziss,$^{3}$ Marvin~A.J.v.~Tilburg,$^{1}$ Chenyang~Mao,$^{1}$ Yizhen~Ren,$^{1}$ Victor~T.v.~Lange,$^{1}$ Sebastian~K\"olling,$^{1,\ddagger}$ Marcel~A.~Verheijen,$^{1,5}$ David~Busse,$^{4}$ Claudia~R\"odl,$^{2}$ J\"urgen~Furthm\"uller,$^{2}$ Friedhem~Bechstedt,$^{2}$ Julian~Stangl,$^{3}$ Johnathan~J.~Finley,$^{4}$ Silvana~Botti,$^{2}$ Jos~E.M.~Haverkort,$^{1}$ Erik~P.A.M.~Bakkers$^{1,\dagger}$
			}
	\bigskip
	\small
	
	$^*$ These authors contributed equally to this work.
	
	$^{\dagger}$ Correspondence to E.P.A.M.B. (\href{mailto:e.p.a.m.bakkers@tue.nl}{e.p.a.m.bakkers@tue.nl})
	
	\vspace*{0.5cm}
\end{center}	

\normalsize
\textbf{This document contains:}\\\\

\textbf{Supplementary Figures:}\\
\textbf{Supplementary Figure~\ref{figs1} $|$} Crystal Quality of the WZ GaAs NW Cores\\
\textbf{Supplementary Figure~\ref{figs2} $|$} Full series of symmetric (0008) reflections of Hex-Si$_{1-x}$Ge$_{x}$\\
\textbf{Supplementary Figure~\ref{figs3} $|$} Schematic Illustration of the nanowires growth process\\
\textbf{Supplementary Figure~\ref{figs4} $|$} Representative large area SEM of Hex-Ge sample \\
\textbf{Supplementary Figure~\ref{figs5} $|$} Atom probe tomography Characterization of Hex-Si$_{0.0.25}$Ge$_{0.75}$\\
\textbf{Supplementary Figure~\ref{figs6} $|$} Comparison between different generations of Hex-Ge samples\\
\textbf{Supplementary Figure~\ref{figs7} $|$} Arrhenius plots of Hex-Si$_{0.20}$Ge$_{0.80}$ with varying quality\\

\textbf{Supplementary Text:}\\
\textbf{Section~\ref{section1} $|$} Reciprocal Space Maps\\
\textbf{Section~\ref{section2} $|$} Fitting using the Lasher-Stern-Würfel model\\
\textbf{Section~\ref{section3} $|$} Temperature dependence of the fundamental bandgap\\
\textbf{Section~\ref{section4} $|$} Temperature dependence of the integrated photoluminescence intensity\\
\textbf{Section~\ref{section5} $|$} Excitation power dependence of the integrated photoluminescence intensity\\
\textbf{Section~\ref{section6} $|$} Temperature dependence of the radiative lifetime\\
\textbf{Section~\ref{section7} $|$} Comparison with group III-V semiconductors \\
\textbf{Section~\ref{section8} $|$} Radiative efficiency and B-coefficient of Hex-SiGe.\\
\textbf{Section~\ref{section9} $|$} Comparison with previous generation Hex-Ge nanowire shells\\

\textbf {References}

\clearpage
\begin{figure}
\centering
\vspace*{2cm}
\includegraphics[width=0.9\columnwidth]{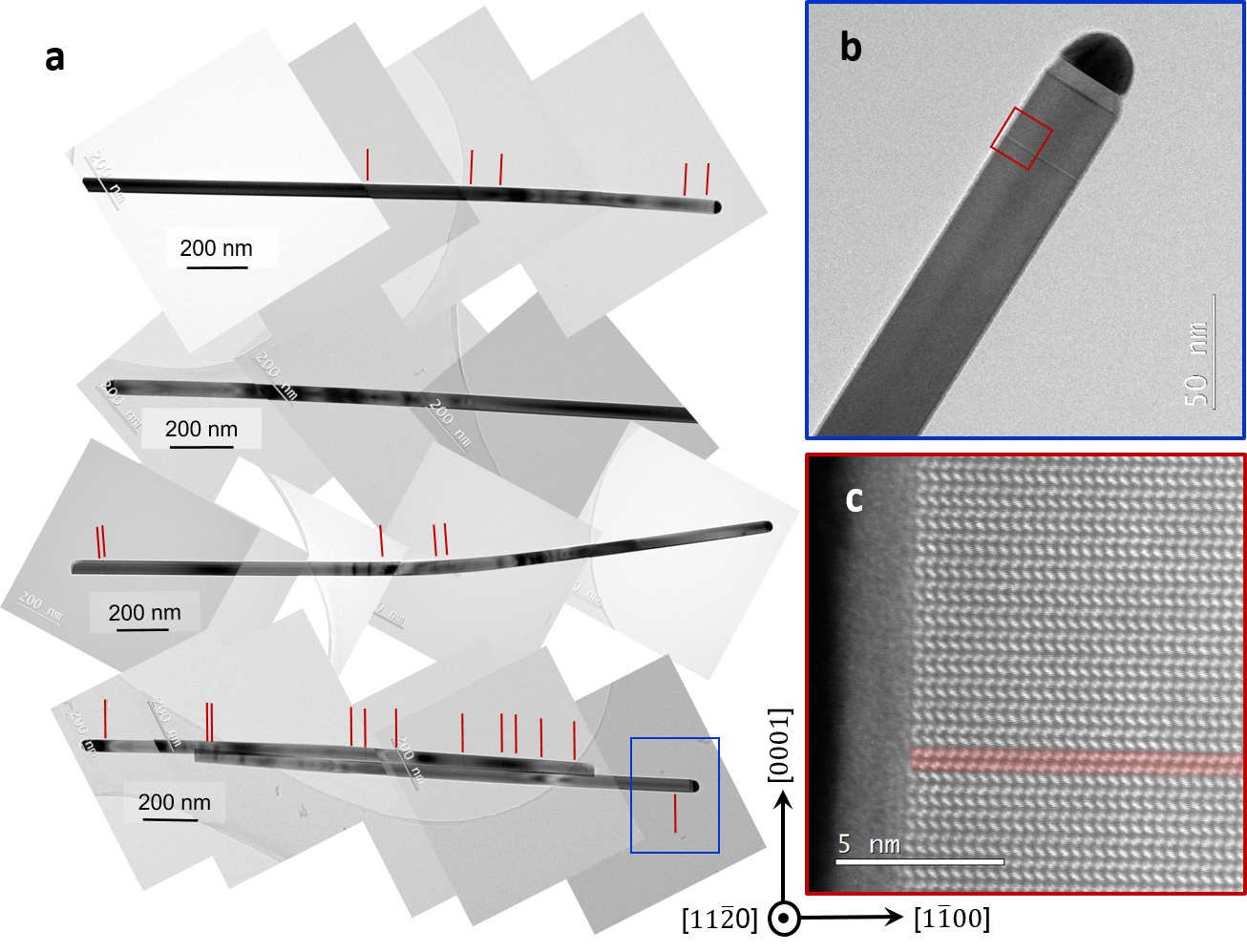}
\caption{\textbf{Crystal quality of the WZ GaAs nanowire Cores:}  \textbf{a,} Bright Field TEM images recorded in the [11$\bar{2}$0] zone axis of 5 representative GaAs core NWs of a pure WZ crystal where stacking faults are indicated with a red line, resulting in a stacking fault density of (0-6\,SFs/$\mu$m).  \textbf{b,} A zoomed in bright field TEM image of the top part of one of the NWs in (a) (highlighted with a blue box) to indicate the purity of the crystal structure.  \textbf{c,} HAADF-STEM image of the highlighted part with a red box in (b) displaying the ABAB stacking of the GaAs atomic columns; the hall mark of the hexagonal crystal structure. The red color highlights a stacking fault forming one cubic layer in the hexagonal structure.}
\label{figs1}
\end{figure}

\clearpage
\begin{figure}
\centering
\vspace*{2cm}
\includegraphics[width=0.8\columnwidth]{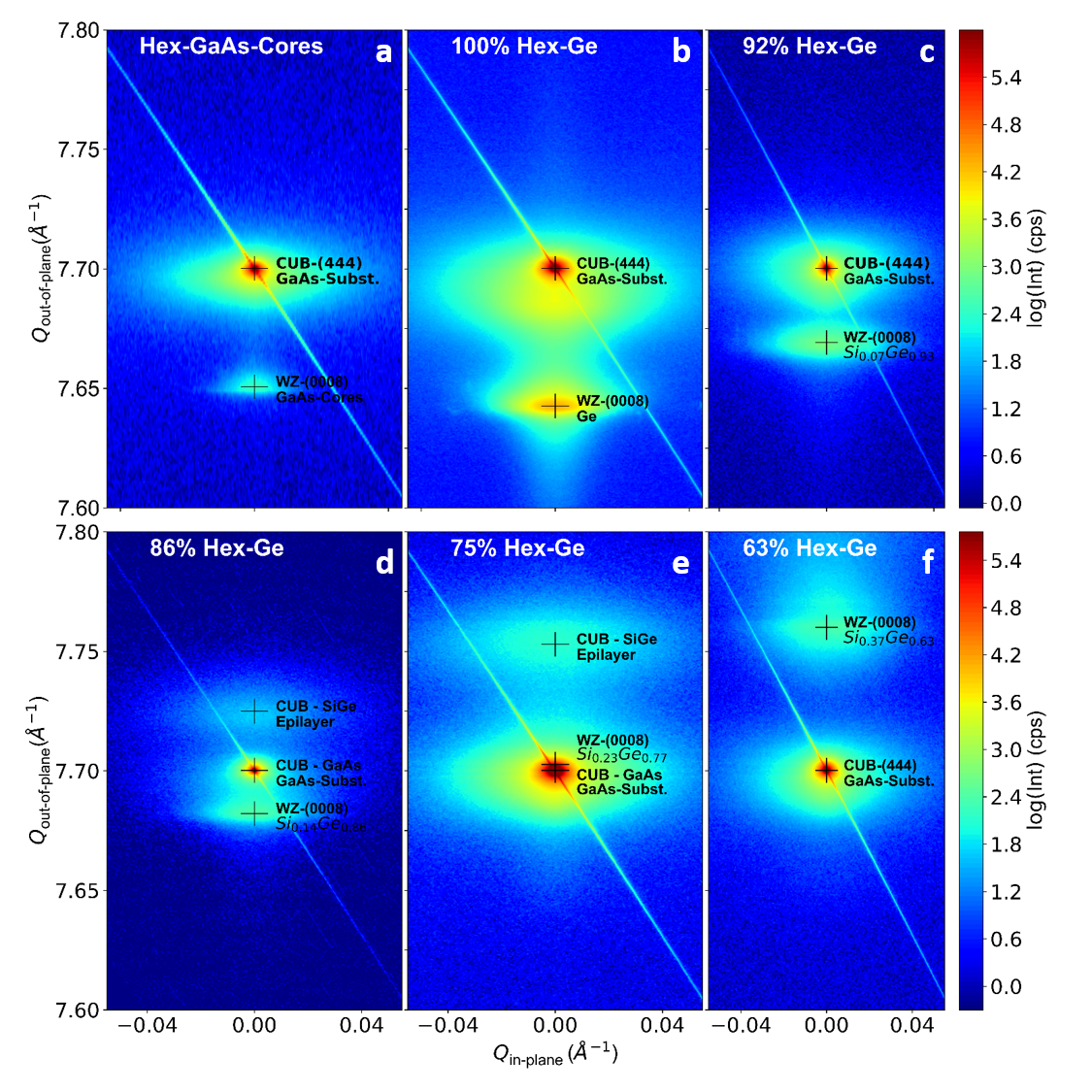}
\caption{\textbf{Full series of symmetric (0008) reflections of Hex-Si1-xGex:}
\textbf{a,} shows a Reciprocal Space Map (RSM) of as-grown WZ GaAs NWs on a cubic GaAs substrate, containing the WZ-GaAs (0008) reflection and the cubic GaAs (444) reflection. \textbf{b,} shows an RSM for a similar sample as in (a) yet with a thick Ge-shell, including the cubic GaAs (444) substrate reflection and the Hex-Ge (0008) reflection. Additional RSMs are shown for samples with Si$_{1-x}$Ge$_x$ shells, in \textbf{c,} ($x = 0.92$), \textbf{d,} ($x = 0.86$), \textbf{e,} ($x = 0.75$),\textbf{f,} ($x=0.63$), as also listed in supplementary Table.~\ref{tab:st1}. A clear increasing shift of Q$_\mathrm{out-of-plane}$ can be observed for increasing Si-content, corresponding to a decreasing lattice constant. For the RSMs in (d) and (e) also a reflection from a parasitic, epitaxial cubic SiGe layer is found.}
\label{figs2}
\end{figure}

\clearpage
\begin{figure}
\centering
\vspace*{2cm}
\includegraphics[height=0.45\textheight]{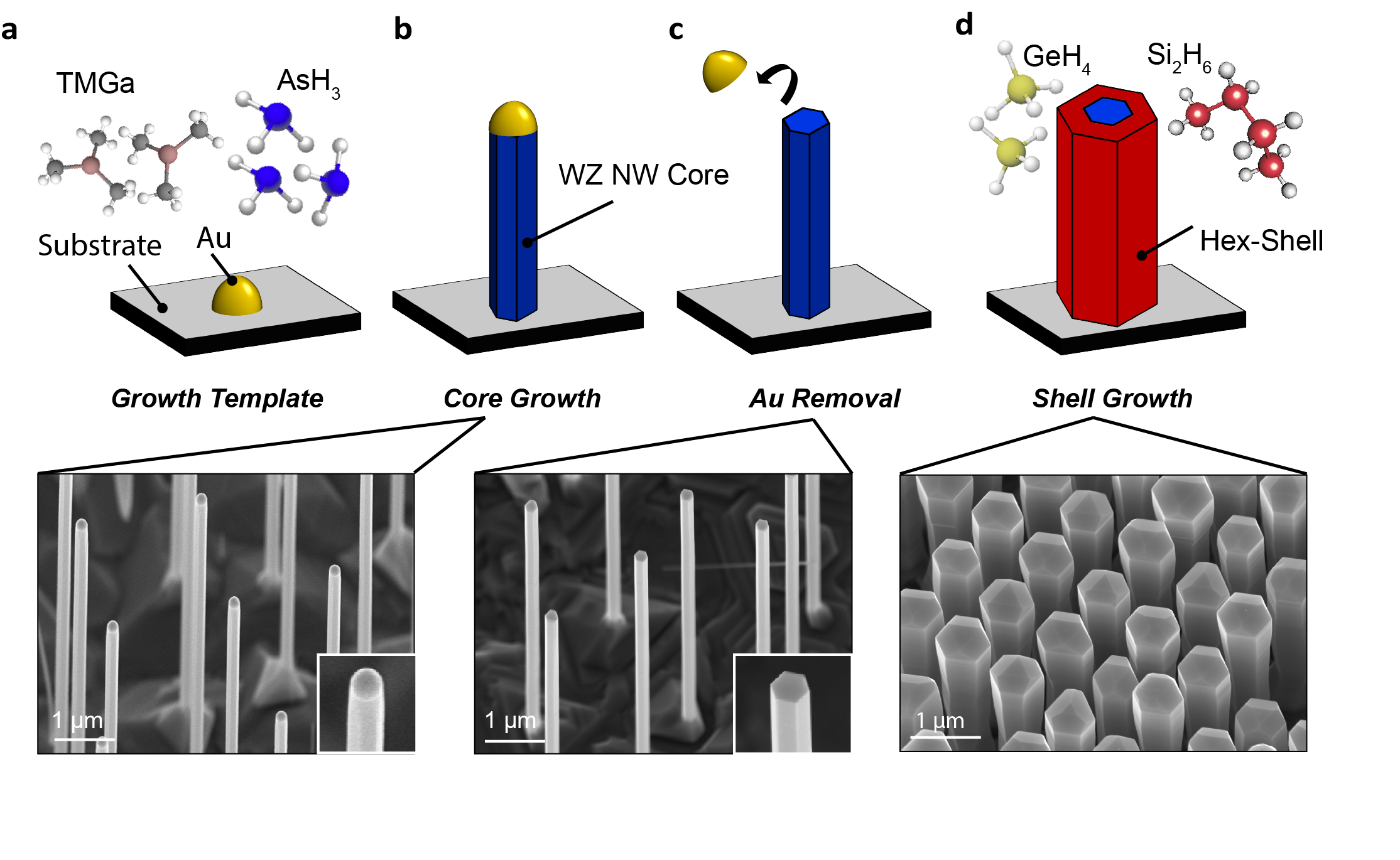}
\caption{\textbf{Schematic illustration of the nanowires growth process:} The core NWs growth starts with \textbf{a,} a substrate patterned with Au catalyst seeds, which is introduced in the MOVPE reactor and annealed at a temperature higher than the eutectic temperature forming an alloy between the catalyst seed and the substrate. \textbf{b,} Afterwards, the GaAs gas precursors (TMGa and AsH$_3$) are introduced, Au-catalysed GaAs core NWs are grown. To proceed with the SiGe shell growth: \textbf{c,} Au seeds are chemically etched away from the GaAs cores, and \textbf{d,} the sample is reintroduced in the MOVPE reactor. A Hex-Si$_{1-x}$Ge$_{x}$ shell is epitaxially grown around the GaAs cores from (Si$_2$H$_6$ and GeH$_4$) precursors. (The molecules are drawn with the freely available MolView Software). The 30\,\textdegree tilted SEM images in the bottom panel are representative to the results of the growth steps in the top panel.}
\label{figs3}
\end{figure}

\clearpage
\begin{figure}
\centering
\vspace*{2cm}
\includegraphics[height=0.4\textheight]{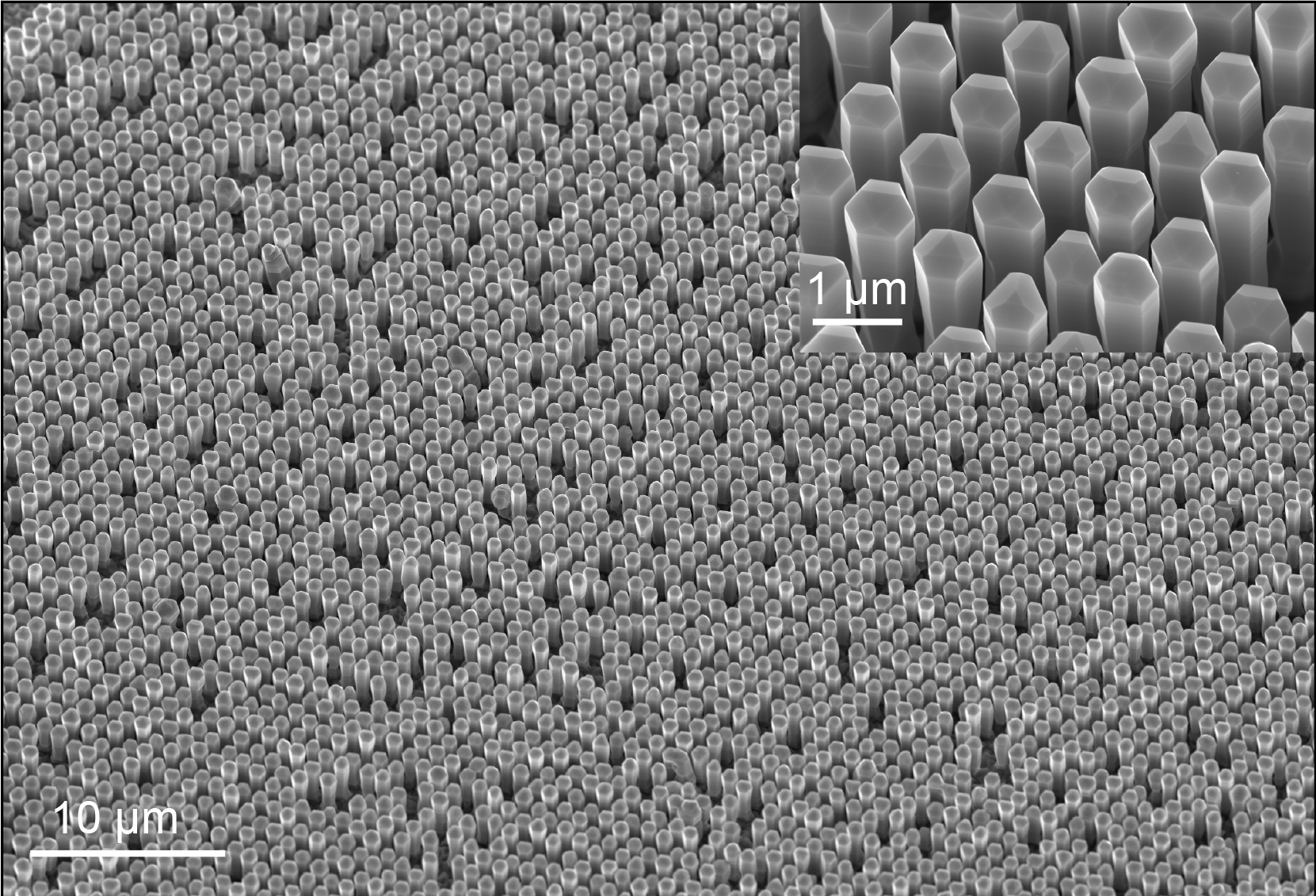}
\caption{\textbf{Representative large are SEM of Hex-Ge sample:} An overview scanning electron microscopy (SEM) image of Hex-Ge/GaAs Core shells showing the uniformity of the growth across the sample with an inset displaying a magnified image of the NW arrays.}
\label{figs4}
\end{figure}

\clearpage
\begin{figure}
\centering
\vspace*{2cm}
\includegraphics[width=0.9\columnwidth]{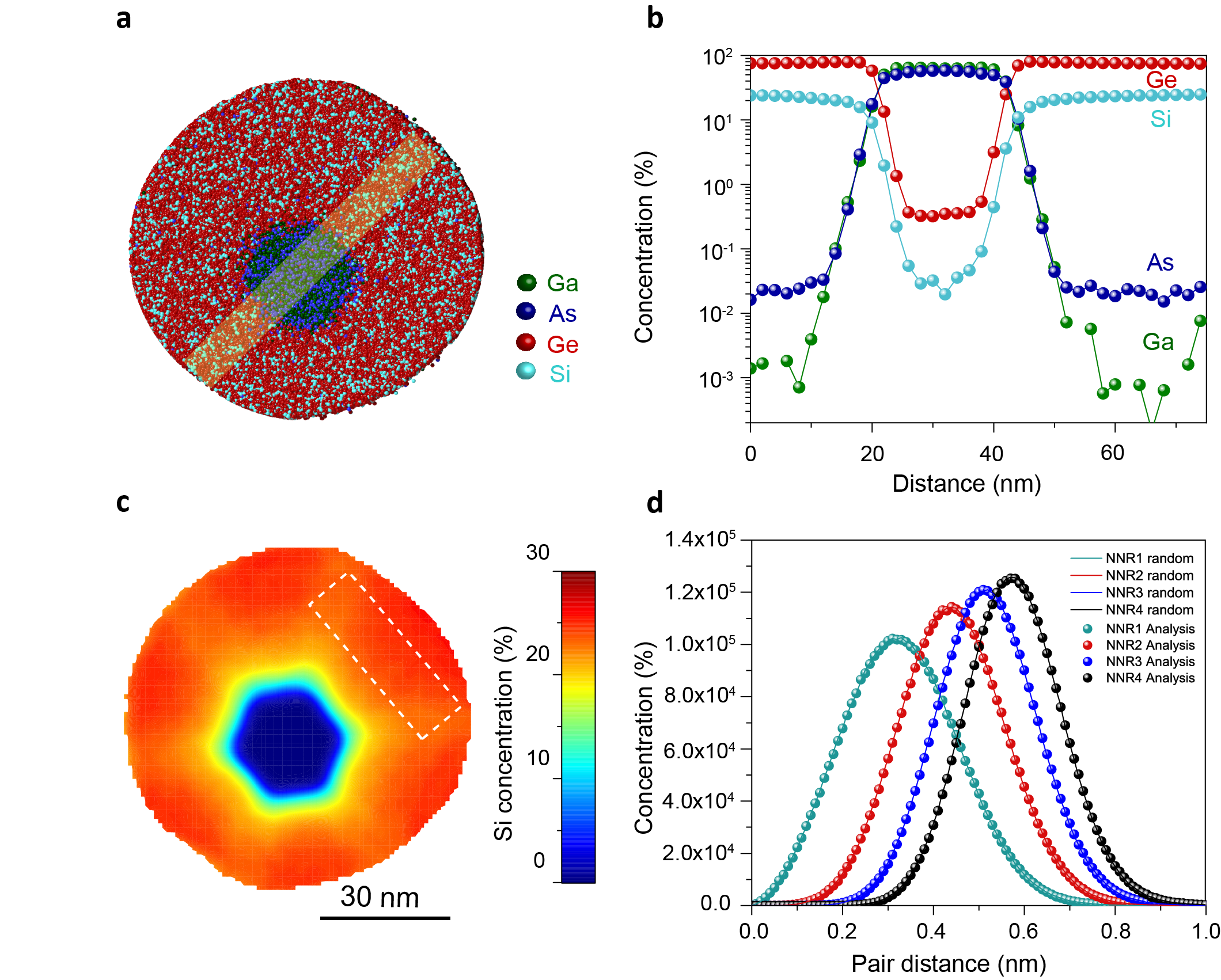}
\caption{\textbf{Atom probe tomography characterization of Hex-Si$_{0.25}$Ge$_{0.75}$:} \textbf{a,} A 3D volume reconstruction of part of a Hex-Si${0.25}$Ge$_{0.75}$ core/shell NW with thicknesses of 35\,nm/46\,nm. For clarity only a slab of 40\,nm thick of the entire 1100\,nm long analyses is shown. Ge (red) and Si (cyan) can clearly be seen to form a shell around the hexagonal Ga (green), and As (blue) core. \textbf{b,} A plot of the atomic species concentration in the SiGe shell in the highlighted yellow rectangle in (a) as a function of the radial distance across the core/shell structure. Every data point in the plot represents a 2\,nm slice taken along the entire length of the nanowire analyses excluding the cubic top part. Constant incorporation of As at a level of approximately 200\,ppm is observed in the entire shell while the Ga concentration quickly drops to a value close to the noise level of ~10\,ppm. \textbf{c,} A radial profile of the SiGe core/shell structure from the APT measurement integrated over a 1.0\,$\mu$m length of the structure showing a Si content of around 25$\%$ as shown in (b). On the highlighted dotted rectangular volume of (c), we carry out a nearest neighbor analysis for Si atoms as previously used to evaluate random alloys of GeSn\cite{Assali,kumar2015}. The nearest neighbor analysis evaluates the distances between each Si atoms pair and its first (to fourth) neighbors. \textbf{d,} A plot comparing the nearest neighbor analysis on the measurement data to a randomized data set. This gives us no indication of Si clustering and has been established as a reliable way to evaluate random alloys\cite{koelling2016impurity}}
\label{figs5}
\end{figure}

\clearpage
\begin{figure}
\centering
\includegraphics[width=0.9\columnwidth]{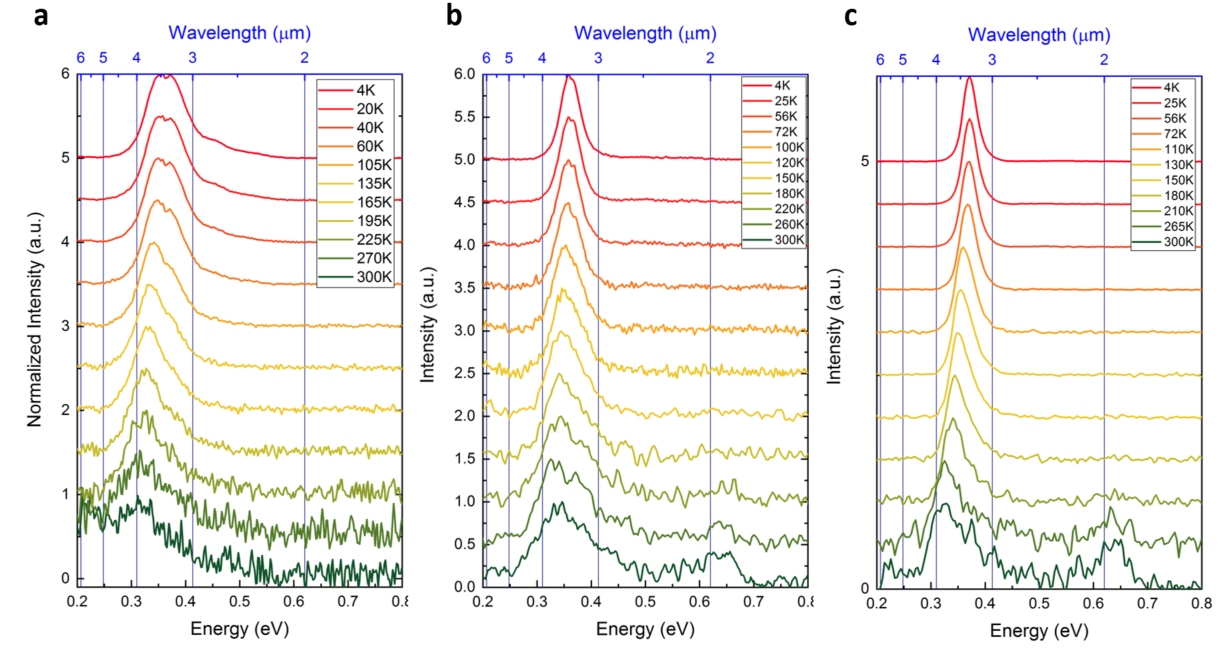}
\caption{\textbf{Comparison between different generations of Hex-Ge samples:}
\textbf{a,} shows the photoluminescence from the first Hex-Ge shell, which was grown using a WZ-GaP core, thus creating many defects due to a large lattice mismatch between the core and the shell. The first Hex-Ge grown on a lattice-matching GaAs shell is shown in \textbf{b,} where the Hex-Ge is grown at a temperature of 600\,\textdegree C. \textbf{c,} shows the spectra of Hex-Ge shells grown at a temperature of 650\,\textdegree C further improving the optical quality.}
\label{figs6}
\end{figure}

\clearpage
\begin{figure}
\centering
\includegraphics[width=0.4\columnwidth]{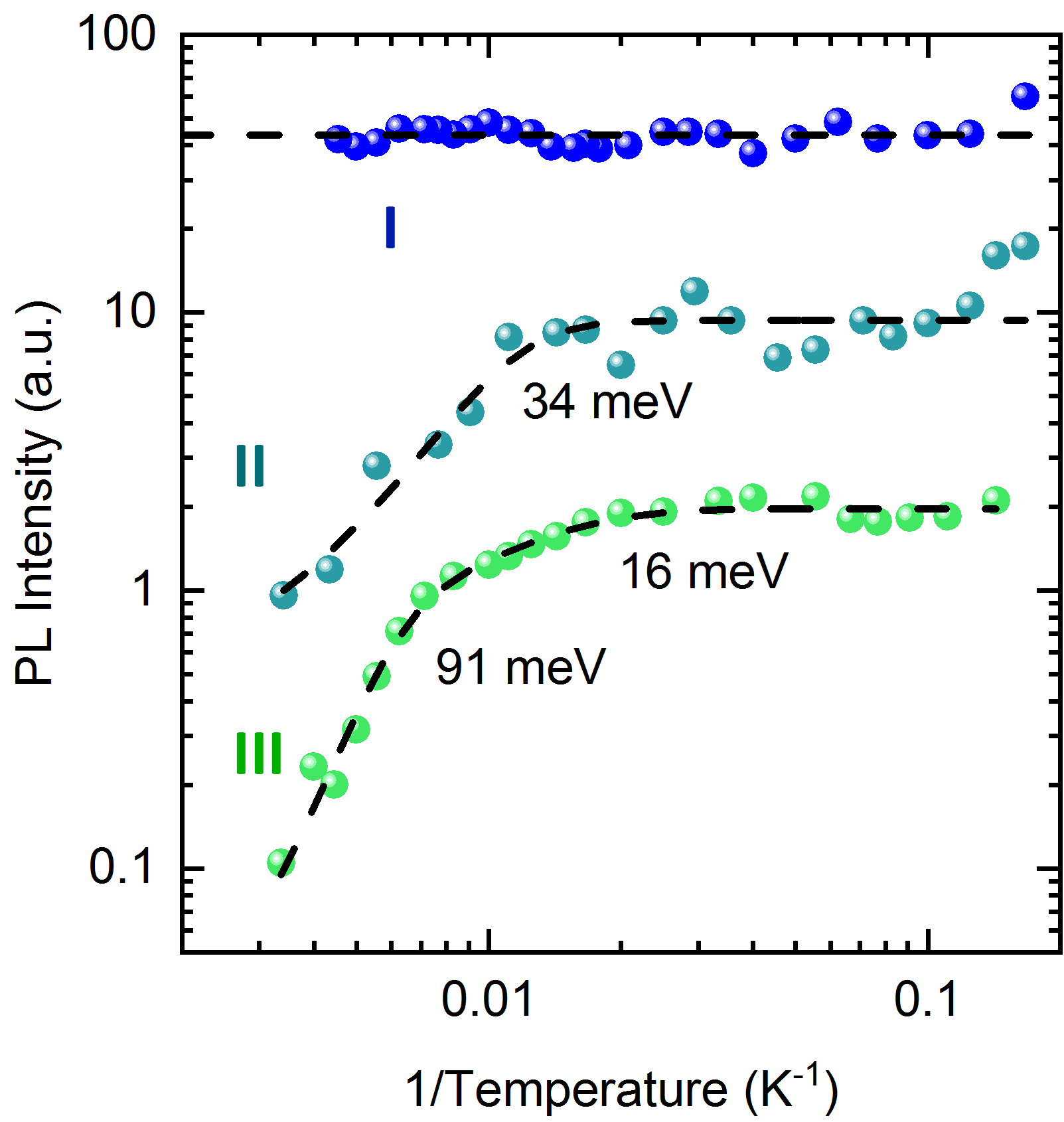}
\caption{\textbf{Arrhenius plots of Hex-Si$_{0.20}$Ge$_{0.80}$ with varying quality:}
 The plots show the same data as presented in Fig.~\ref{fig4}c, but here presented in an Arrhenius representation. For the lowest quality sample III, two non-radiative processes are found with activation energies of 16\,meV and 91\,meV. For sample II only a single activation energy is found of 34\,meV where sample I does not show any decay in intensity over the full measured temperature range. Details of the samples are given in supplementary Table.~\ref{tab:st2}.}
 \label{figs7}
\end{figure}

\clearpage
\noindent

\section{Reciprocal Space Maps} \label{section1}
For all measured samples at least 3 individual hexagonal reflections have been measured. For the pure Hex-Ge sample, the azimuth was varied to enhance the fidelity of the extracted lattice parameters. In addition, a cubic GaAs substrate reflection has always been used as an anchor in reciprocal space to correct for any possible alignment offsets. From the measured symmetric reflections, see the full series in supplementary Fig.~\ref{figs2}. One could calculated and correct for the tilt of the lattice-planes ([111], [0001]) with respect to the samples surface. Furthermore, the Q$_\mathrm{out-of-plane}$ position of the WZ (0008) reflection allows to calculate the $c$-lattice parameter, corresponding to the [0001]-crystal direction. For some Ge-concentration the (0008) NW reflection coincides with the cubic (444) substrate refection (see Supplementary Fig.~\ref{figs2}) which makes a systematic evaluation of the symmetric RSMs complicated, hence also asymmetric space maps, around reflections that are only allowed in the hexagonal crystal-lattice have been measured. The position of the asymmetric reflections in reciprocal space allows to extract the in- as well as the out-of-plane lattice parameters ($a$, $c$). In Fig.~\ref{fig2}, a series of Hex-(1$\bar{0}$18) reflections for all measured Ge-concentrations is shown, the peak-position sensitively depends notably on the amount of Ge present in the alloy, lower Ge-concentrations result in lower lattice parameters ($a$ and $c$), which are closer to the native lattice parameters of Hex-Si\cite{Hauge2017}. For all the shown RSMs the Q$_\mathrm{out-of-plane}$ direction corresponds to the crystalline [0001] direction, and the Q$_\mathrm{in-plane}$-direction corresponds to the [1$\bar{0}$10] direction, both in the hexagonal system, indicated by the four Miller-Bravais indices.\\
\\
To accurately determine the peak positions, all RSMs were corrected according to the peak positions of the cubic GaAs-substrate reflections to eliminate any angular alignment offsets. Then a 2D-Gauss fit was performed on the data-sets in q-space before gridding, to reduce the influence of possible artefacts coming from the gridding-routine. For plotting the dataset, the irregularly spaced q-coordinates, as measured and transformed from the angular-space, have been gridded into a regularly spaced q-coordinate system.
The combined results from the XRD measurements can be found in Supplementary Table.~\ref{tab:st1} where the measured lattice parameters are given for each measured Ge-concentration. For all samples the influence of the WZ-GaAs core material on the Si$_{1-x}$Ge$_{x}$ lattice parameter can be neglected because of the fact that a relatively thin GaAs core (around 35\,nm) is surrounded by a thick (several 100\,nm) Si$_{1-x}$Ge$_{x}$ shell. Hence, the crystalline properties of the Hex-Si$_{1-x}$Ge$_{x}$ shell dominate the whole structure. Furthermore, Hex-Ge and WZ-GaAs are nearly lattice matched (see lattice parameter of WZ-GaAs\cite{Jacobsson2015} which implies that basically no strain in the shell is expected for the samples with high Ge-concentrations ($>60\%$) as also confirmed by FEM-simulation. This is an important aspect since it confirms the high fidelity of the found lattice parameters, especially for the lattice parameter of pure Hex-Ge.\\
\\
The errors given in supplementary Table.~\ref{tab:st1} consider the accuracy of defining the peak position with a 2D-fit as described, as well as the scattering of the individual lattice parameter values extracted from the evaluation of multiple peaks. The instrumental resolution can be neglected for the error estimation, since the contribution to the given errors will be much smaller than the total error-values.

\begin{table}[ht]
\caption{\label{tab:st1} Hexagonal in-plane ($a$) and out-of-plane ($c$) lattice parameters of all measured Hex-Si$_\mathrm{1-x}$Ge$_\mathrm{x}$ samples with corresponding error-values extracted from XRD measurements.}
\begin{ruledtabular}
\begin{tabular}{lcc}
Ge-content & a\,(\AA) &b\,(\AA)  \\ \hline
1.00&	$3.9855\pm0.0003$&	$6.5772\pm0.0003$\\
0.92&	$3.9789\pm0.0001$&	$6.5542\pm0.0001$\\
0.86&	$3.9649\pm0.0005$&	$6.5431\pm0.0004$\\
0.75&	$3.9505\pm 0.0008$&	$6.5257\pm0.0001$\\
0.63&	$3.9206\pm0.0000$&	$6.4790\pm0.0005$\\
\end{tabular}
\end{ruledtabular}
\end{table}

\section{Fitting using the Lasher-Stern-Würfel model}\label{section2}
The observed photoluminescence spectra of Hex-Ge and Hex-SiGe all consist out of a single peak. We attribute the observation of a single photoluminescence peak to a band-to-band (BtB) recombination. The absence of excitonic effects at low temperatures is due to an As-doping level of $9\times10^{18}$\,cm$^{-3}$ as deduced by Atom Probe Tomography shown in Supplementary Fig.~\ref{figs5}. At this doping level, individual As-dopants, which are expected to be shallow in a small bandgap semiconductor, will merge into a doping band which at its turn is expected to merge\cite{Alexander1968} with the conduction band. GaAs NWs with a similar doping level\cite{Chen2017} also show single peak photoluminescence spectra which are very similar to our findings in Hex-SiGe. To accurately establish whether the observed photoluminescence is due to BtB recombination, we fitted the experimental spectra with the Lasher-Stern-Würfel (LSW) model\cite{Chen2017,katahara2014,P.Wurfel1982,Lasher1964}. This model, that predicts the shape of a photoluminescence peak, is derived from the Planck-Einstein radiation law\cite{Einstein1917} and is given by:
\begin{equation} \label{eq:IPL}
   I_\mathrm{PL} = \frac{2\pi}{h^3c^2} \frac{E^2a(E)}{\exp{\left(\frac{E-\Delta\mu}{KT}\right)}-1}
\end{equation}
In this equation $\Delta\mu$ is the splitting of the quasi-fermi levels and $a(E)$ is the absorptivity. In modelling the absorptivity, parabolic bands have been assumed. Corrections for an Urbach tail and an excitation dependent Burstein-Moss shift have been made in analogy with Katahara et al.\cite{katahara2014}. We have fitted both the temperature dependent and the excitation power dependent photoluminescence measurements as shown in Fig.~\ref{fig3}a . The high quality fits by the LSW model unambiguously show that the observed photoluminescence is exclusively due to BtB recombination. It is of paramount importance for the analysis performed in the main text that measured recombination lifetimes are due to BtB recombination and not due to e.g. an impurity or defect related optical transition. We note that the deduced carrier temperature exceeds 700\,K at the highest excitation densities. A detailed analysis of the fitting procedure including a detailed analysis of the observed Burstein-Moss shift\cite{Moss1954} and the observed carrier temperature is beyond the scope of the present paper.

\section{Temperature dependence of the fundamental bandgap}\label{section3}

Although the temperature dependence of the fundamental bandgap is most often described by the Varshni equation\cite{Varshni1967}, the Vina equation\cite{Vina1984} provides a more accurate description for elevated temperatures
\begin{equation} \label{eq:Eg}
  E_\mathrm{g} = a-b\left(1+\frac{2}{exp{(\frac{\theta}{T})}-1}\right)
\end{equation}
in which $a$ is a constant, $b$ represents the strength of the electron-phonon interaction, and $\theta$ is the Debye temperature of the material. For the bandgap of Hex-Ge the Vina equation is fitted in Fig.~\ref{fig3}c where the following values are found; $a=0.36$\,eV, $b=9.2$\,meV and a Debye temperature of $\theta=66$\,K. The shrinkage of the Si$_\mathrm{0.20}$Ge$_\mathrm{0.80}$ bandgap, which is displayed in Fig.~\ref{fig3}c follows a different behavior due to compositional fluctuations\cite{polimeni2000effect} of the crystal. The initial fast shift of the apparent bandgap is probably due to the carrier thermalization towards compositional pockets with lower bandgap while the apparent deviation from the Vina law at high temperature is most probably due to the fact that the spectrum should be interpreted as a convolution of the Fermi-Dirac distribution with a Gaussian broadening function due to the compositional fluctuations, the details of which are beyond the scope of the present paper.

\section{Temperature dependence of the integrated photoluminescence intensity.}\label{section4}
Here, we provide a detailed Arrhenius analysis of the temperature dependence of the integrated PL as presented in Fig.~\ref{fig4}c. Our goal is to provide quantitative information about the ratio between the radiative and non-radiative recombination rates. In order to explain the temperature dependence of the photoluminescence emission intensity, we first have to set up the proper rate equation model. Since the donors have merged into a donor band which shifted into the conduction band, we will not incorporate the donor level into the rate equation model. Following the LSW analysis in Section~\ref{section2}, we concluded that the photoluminescence spectrum can be explained by Band-to-Band (BtB) recombination with only a minor influence of the acceptor related transition. As a consequence, we will limit our rate equation model to a three level system incorporating the conduction band, the valence band and a “killer defect” which is characterized by an activated non-radiative recombination lifetime. We thus use the one-center model in the classification of Reshchikov\cite{Reshchikov2014}, which is explained in more detail by a configuration coordinate diagram. In this one-center model, the internal quantum efficiency ($\eta_\mathrm{int}$) for radiative emission varies with temperature according to the ratio of the radiative recombination rate, divided by the total recombination rate by $ \eta_\mathrm{int} = \frac{\tau_r^{-1}}{\tau_\mathrm{r}^{-1}+\tau_\mathrm{nr}^{-1}(T)}$. The low excitation data collected at 68\,W/cm$^2$, which are presented in Fig.~\ref{fig3}, can be fitted with this formula by assuming that the non-radiative recombination rate is thermally activated by $\tau_\mathrm{nr}^{-1}(T) = \tau_\mathrm{nr,o}^{-1} e^{-E_a/KT} $ similar to III-V materials\cite{Leroux1999,BacherH1992,Lourenco2003,Schenk2000,Lambkin1994} 
The excellent quality of the Arrhenius fit provides evidence that the non-radiative recombination into the yet unknown killer defect can indeed be explained by an activated non-radiative recombination rate. The temperature dependence of the photoluminescence intensity can thus be expressed as
\begin{equation} \label{eq:I_T1}
   I(T) = \frac{I_0}{1+R_Ae^{\frac{-EA}{KT}}}
\end{equation}
in which the photoluminescence quenching rate\cite{Leroux1999,BacherH1992,Lourenco2003,Schenk2000,Lambkin1994} into the non-radiative centre is given by $R_A=\frac{\tau_r}{\tau_(nr,A,0)}$ . In most semiconductors, different non-radiative recombination centres exist which feature e.g. activation energy $E_A$, $E_B$ and quenching rates $R_A$, $R_B$ resulting in Eq.~\ref{eq:I_T2}:
\begin{equation} \label{eq:I_T2}
   I(T) = \frac{I_0}{1+R_Ae^{\frac{-E_A}{KT}}+R_Be^{\frac{-E_B}{KT}}}
\end{equation}
It is instructive to perform this analysis to three different generations of Hex-SiGe samples which are specified in supplementary Table.~\ref{tab:st2} and whose Arrhenius plots are shown in supplementary Fig.~\ref{figs7}. In sample III, we observe a first quenching mechanism with activation energy $E_A=16±1$\,meV with a quenching efficiency of $R_A=3\pm1$, and a second quenching mechanism with $E_B=91\pm2$\,meV and $R_B=6\times10^2\pm±1$, which is at least partially due to surface recombination. These rates imply an internal quantum efficiency of  $\frac{\tau_r^{-1}}{\tau_{r}^{-1}+\tau_{nr,A}^{-1}+\tau_{nr,B}^{-1}} = 0.15\%$  when both non-radiative channels are fully activated (room temperature). The first quenching mechanism seems to have disappeared in sample II which was grown at a higher temperature. In sample II, we only observe photoluminescence quenching above a temperature of 100\,K, which is again tentatively attributed to be at least partially due to surface recombination. The activation energy $E_B=34\pm5$\,meV is tentatively explained by the de-trapping from localized states due to alloy fluctuations in the Hex-SiGe nanowire shell. Once the carriers are de-trapped, they will quickly diffuse to the nanowire surface where they recombine non-radiatively. In sample I, both quenching mechanisms have disappeared as $R_A=\frac{\tau_r}{\tau_{nr,A,0}} =0$ and $R_B=\frac{\tau_r}{\tau_{nr,B,0}} =0$ at an excitation density of 36\,kW/cm$^2$, thus showing that sample I remains in the radiative limit up to 220\,K. The quality of sample I is probably higher due its thick Hex-SiGe shell which reduces the amount of surface recombination as well as by its length which reduces the influence of re-evaporating arsenic (As) and gallium (Ga) from unwanted growth on the substrate. To be completely sure, we have regrown sample I resulting in an identical temperature dependence as the first grown sample.\\
\begin{table}
\caption{Growth parameters of the three studied different Hex-SiGe samples with increasing quality and the dimensions of the NWs presented in Fig.~\ref{fig4}b and ~\ref{fig4}c.}
\begin{ruledtabular}
\begin{tabular}{cccccc}
Sample$\#$&	Growth Temp\,(\textdegree C)&	Ge-content($\%$)&	NW Shell Radius\,(nm)&	NW Core Diameter\,(nm)&	NW Length\,($\mu$m)\\ \hline
Sample I&	700&	79&	    650& 	175& 	8\\
Sample II&	700&	80&	    400&	35&  	2.5\\
Sample III&	650&	75&	    150& 	35&  	2.5\\
\end{tabular}
\end{ruledtabular}
\label{tab:st2} 
\end{table}

\section{Excitation power dependence of the integrated photoluminescence intensity.}\label{section5}
At low excitation density, $\Delta n<n_0$, the nonradiative, radiative and Auger processes all yield a linear dependence of the PL-intensity versus excitation power with a slope of unity, which suggest that we are not allowed to draw any conclusions from the data in this range of excitation power. However, this simplified analysis assumes that the non-radiative recombination centres are not being saturated. Since we do not observe any deviation from a linear behavior, our data suggest that, even if non-radiative recombination centres would be present, we are not able to saturate them with excitation power. This suggests that we do not have any non-radiative recombination centres in the bulk of the material, implying that we are already in the radiative limit. We note that this argument applies both for  $\Delta n<n_0$ and $\Delta n>n_0$.\\
\\
At high excitation density, $\Delta n>n_0$, we will use the analysis of Yoo et al.\cite{Yoo2013}. In their analysis, the total carrier generation rate $G$ should be equal to the total carrier recombination rate by 
\begin{equation} \label{eq:G1}
  G = An+Bn^2+Cn^3
\end{equation}
in which An is the Shockley-Read-Hall nonradiative recombination rate, $Bn^2$ is the radiative recombination rate and $Cn^3$ is the Auger nonradiative rate. At high excitation density (which is above 500\,W/cm$^2$ for Hex-Ge as shown by bandfilling in Fig.~\ref{fig3}a) when the photo-injected carrier densities $\Delta n$, $\Delta p$ are larger than the electron concentration due to unintentional As-doping (see APT measurements in Supplementary Fig.~\ref{figs5}), we expect the behavior as predicted by Yoo et al.\cite{Yoo2013}.
\begin{equation} \label{eq:G2}
G = A\sqrt{\frac{I_{PL}}{aB}}+\frac{I_{PL}}{a}+C\left(\frac{I_{PL}}{aB} \right)^{3/2}
\end{equation}

In the plot of the integrated photoluminescence intensity versus excitation density\cite{Schmidt1992}, Eq.~\ref{eq:I_T2} yields a slope of two for non-radiative recombination (provided that the non-radiative recombination centres are not being saturated, see above), a slope of unity for radiative recombination and a slope of 2/3 for Auger recombination. We note that we do not observe a decrease of the PL-intensity at the highest excitation power, providing a first indication that Auger recombination losses are not yet dominant in this material.\\
\\
For the Hex-Si$_{0.20}$Ge$_{0.80}$ sample, we are not yet able to establish a clear boundary between the $\Delta n<n_0$ and the $\Delta n>n_0$ regime due to the added complication of alloy broadening. Most probably, the Si$_{0.20}$Ge$_{0.80}$  alloy will be composed out of compositional pockets in which either $\Delta n<n_0$ or $\Delta n>n_0$ applies. The observation of a slope of exactly unity, as shown in the inset of Fig.~\ref{fig4}c, implies that both type of pockets are in the radiative limit.

\section{Temperature dependence of the radiative lifetime}\label{section6}
As shown in Fig.~\ref{fig4}b, we observe a temperature independent recombination lifetime in sample I. In this section, we will show that such a T-independent recombination lifetime can only be explained by radiative recombination in a degenerately doped semiconductor.Non-radiative recombination features an activated behavior at low temperature which is governed by $\tau_\mathrm{nr}^{-1}(T) = \tau_\mathrm{nr,o}^{-1} e^{-E_a/KT}$ as explained in Section~\ref{section4}. By analyzing the well-known expressions for the Shockley-Read-Hall (SRH) non-radiative recombination mechanism for intrinsic material, the SRH lifetime can be expressed\cite{Wirths2015,Schubert2006} as in $\tau_{SRH}=\tau_{p,0}\left(1+cosh\frac{E_t-E_i}{K_BT} \right)$ which $E_t$ is the trapping level, $E_i$ is the intrinsic Fermi level and $\tau_{p,0}$ is the SRH lifetime for minority holes. At higher temperature, the SRH lifetime is expected to decrease with $T^{-1/2}$ due to the fact that both $\tau_{n0}$ and $\tau_{p0}$ are inversely proportional to the thermal velocity. We conclude that it is clearly not possible to interpret the observed temperature independent recombination lifetimes as being due to non-radiative recombination.\\
\\
We subsequently like to discuss Auger recombination which might be expected due to the high $n$-doping by unintentional arsenic (As) incorporation during growth. The Auger rate includes two different processes\cite{Benz1977,Galler2013}, the $nnp$-Auger process in which the excess energy is transferred to an electron and the $npp$-Auger process in which the excess energy is transferred to a hole. In our case, we have high $n$-doping due to As incorporation during growth, resulting in a doping concentration $n_0$. We expect that the $nnp$-Auger process will be most important in our $n$-doped Hex-SiGe samples. The Auger coefficients are however temperature dependent\cite{A.Haug1978}, which results in a T-dependent recombination lifetime, which is not consistent with our observations. Most importantly, as shown in the inset of Fig.~\ref{fig4}c, we observe a linear relation between the integrated photoluminescence intensity and the excitation power. We do not observe a decrease of the PL-intensity at high excitation, which is a strong indication that Auger processes are still weak at our experimental conditions.\\
\\
We are thus left with radiative recombination. The radiative lifetime for an intrinsic semiconductors increases\cite{tHooft1985} with $T^{3/2}$ showing sub-nanosecond radiative lifetimes at low temperature which increase to more than a microsecond at room temperature. For a degenerately doped semiconductor, the radiative lifetime is expected to be temperature independent since the B-coefficient for radiative recombination is proportional to $B\equiv \frac{L}{n_p}$ , in which $L$ is the spontaneous radiative recombination rate\cite{Lasher1964}. It can be easily seen that for a degenerate semiconductor $p\propto T^{3/2}$, $L\propto T^{3/2}$ while $n$ becomes temperature independent. Both the B-coefficient for radiative recombination rate and the radiative lifetime are thus expected to be independent of temperature.\\
\\
We present the photoluminescence lifetime measurements for all three samples in Fig.~\ref{fig4}b. We recall our conclusion from Fig.~\ref{fig4}c that sample III, II and I are in the radiative limit up to 40\,K, 100\,K and $>220\,K$. This behavior is exactly reproduced in Fig.~\ref{fig4}b in which the lifetimes are constant up to 40\,K, 90\,K and $>220\,K$ which is indeed expected for a degenerate semiconductor in the radiative limit. In sample III, II, non-radiative recombination becomes the dominant recombination mechanism above 40\,K and 90\,K, respectively, as is clear from the observed slope which is close to -0.50 as expected for non-radiative SRH recombination at high temperature. The non-radiative recombination at high temperature is expected to be due to recombination at the nanowire surface. In order to again obtain the correct statistics, we performed photoluminescence lifetime measurements on more than 60 different NWs taken from sample I at 4\,K and at 300\,K. The data are displayed in Fig.~\ref{fig4}d. We observe a spread in the recombination lifetimes at 4\,K, which we attribute to variations of the local density of photonic states around each individual wire, the analysis of which is beyond the scope of the present paper.

\section{Comparison with group III-V semiconductors}\label{section7}
The measured lifetime of Hex-Si$_{0.20}$Ge$_{0.80}$ at low temperature is very comparable to the reported recombination lifetimes\cite{Feldmann1987,PtHooft1987,Bellessa1998} in literature for III-V compound semiconductors, which are generally of the order of 1\,ns. Jiang et al.\cite{Jiang2012} reported a temperature independent lifetime of 1\,ns in core/shell GaAs/AlGaAs NWs, very similar to our yet unpassivated Hex-Si$_{0.20}$Ge$_{0.80}$ nanowire shells. The comparison of the quenching ratio of the integrated photoluminescence intensity when increasing the temperature from 4\,K to 300\,K, compares quite favorable for Hex-SiGe where this ratio varies between a factors of 15-100 as shown in Fig.~\ref{fig3}c. Lambkin et al. found a photoluminescence quenching ratio of the order of 105 for InGaP\cite{Lambkin1994}. Louren\c{c}o et al. observed a quenching ratio around 200 for GaAsSbN/GaAs quantum wells\cite{Lourenco2003}. Leroux et al. also observed quenching rations above between 100 for undoped GaN up to 1000 for Mg-doped GaN\cite{Leroux1999}. The PL quenching in Ge microstrips as obtained by Virgilio et al.\cite{Virgilio2015} are comparable to ours.

\section{Radiative efficiency and B-coefficient of Hex-SiGe}\label{section8}
In order to compare the radiative emission strength of Hex-SiGe with other well-known direct bandgap semiconductors like e.g. GaAs or InP, we compare the radiative emission rate at room temperature which is most relevant for device applications. By making the comparison at 300\,K, excitonic effects as well as effects due to carrier localization in the Hex-SiGe alloy are not relevant anymore. The key parameter to compare the radiative efficiency of a semiconductor is the B-coefficient which is a recombination rate, corrected for the doping density.\\
\\
The radiative rate per volume of a semiconductor $R_{rad}$ can be expressed in terms of the B-coefficient, n- and p-type doping concentration $n_0$ and $p_0$ and the number of excited electron-hole pairs $\Delta n=\Delta p$. For a highly $n$-doped semiconductor, which yields $n_0\gg \Delta n$ $R_{rad}$ can be expressed as:
\begin{equation} \label{eq:Rrad}
 R_{rad}=B_{rad}(n_0+\Delta n)(p_0+\Delta p)\approx B_{rad}n_0 \Delta p
\end{equation}
The experimentally observed radiative lifetime $\tau_{rad}$ is determined by the recombination rate per volume $R_{rad}$ and the number of excited electron hole pairs $\Delta n=\Delta p$ such that $\tau_{rad}=\Delta p/R_{rad}$. Combining this result with Eq.~\ref{eq:Rrad} gives a definition for the B-coefficient of:
\begin{equation} \label{eq:Brad}
 B_{rad}=\frac{1}{\tau_{rad}}n_0
\end{equation}
In which $\tau_{rad}$ is the radiative lifetime at 300\,K and $n_0$ is the activated donor density. To determine the $B_{rad}$ coefficient we carefully evaluate the determined values for $\tau_{rad}$ and the doping density $n_0$. The measured photoluminescence-lifetimes show a spread over different wires as shown in Fig.~\ref{fig4}d. We attribute this spread to a variation of the optical density of states of each individual wire. Using the decay times measured at 4\,K (Fig.~\ref{fig4}d) and extrapolating them to 300\,K assuming temperature independence, we deduce an upper limit of 1.6\,ns, while the lower limit is close to 0.4\,ns, as shown by the 300\,K measurements in Fig.~\ref{fig4}d. Because it is of key importance that the measured lifetime is the radiative lifetime we carefully address this point below.

\begin{itemize}
  \item	    One might argue whether the measured photoluminescence decay time at 300\,K, is equal to the radiative lifetime. Our main argument is provided by Fig.~\ref{fig4}e, which shows that the photoluminescence intensity at 300\,K is almost equal to the photoluminescence intensity at 4\,K. Since we know that Hex-SiGe is in the radiative limit at 4\,K, and we observe almost the same emission intensity at 300\,K, it is very clear that Hex-SiGe should remain very close to the radiative limit at 300\,K.
  \item	    A second point of concern might be whether we are still in the degenerate limit $\Delta n<n_0$. The main evidence for this point is that, for most wires, we measure an excitation power independent photoluminescence decay time in the same excitation range as in the inset of Fig.~\ref{fig4}c. In addition, we measure a temperature independent photoluminescence lifetime in Fig.~\ref{fig4}b (see section~\ref{section5}) which can only be understood for a semiconductor with degenerate doping.
\end{itemize}

The donor density $n_0$ has been estimated using two techniques, the first of which is atom probe tomography, shown in Supplementary Fig.~\ref{figs5}, where a donor concentration of $n_0=9\times10^{18}$\,cm$^{-3}$ is found. However this number might differ from the number of active dopants. The active doping concentration can be calculated from the electron-quasi-fermi-level (eQFL) and the density of states in the conduction band. We find an eQFL of 35\,meV using the results of the LSW fitting model as presented in Section~\ref{section2} and the density of states are calculated using the effective masses following DFT calculations 5. Using these values we find a doping level of $n_0=2.3\times10^{18}$\,cm$^{-3}$.\\
\\
Now combining the upper bound for the donor density of $9\times 10^{18}$\,cm$^{-3}$ with the upper bound of 1.6\,ns for the radiative lifetime, we obtain a lower bound for the B-coefficient of $0.7\times10^{-10}$\,cm$^3$/s , which is roughly 2$\times$ smaller than the B-coefficient of InP. Using the lower limits for $n_0$ and $\tau_{rad}$ an upper limit of $11\times10^{-10}$\,cm$^3$/s is found for the B-coefficient, which is 9$\times$ larger as for InP. A comparison of B-coefficients of different III-V materials, cubic Si and Hex-Si$_{0.20}$Ge$_{0.80}$ is made in Supplementary Table.~\ref{tab:st3}. Extracting the B-coefficient and thus the transition matrix elements is of great importance for possible device applications of Hex-SiGe for e.g. lasers, modulators, detectors and LEDs which all critically depend on the strength of the light-matter interaction.\\

\begin{table}[ht]
\caption{\label{tab:st3} Radiative coefficient (B-coefficient) of Hex-Si$_\mathrm{0.20}$Ge$_\mathrm{0.80}$ as calculated in section~\ref{section8} and literature values for GaAs\cite{Lush2009}, InP\cite{Semyonov2010} and cubic Si\cite{Trupke2003}}
\begin{ruledtabular}
\begin{tabular}{cccc}
Hex-Si$_\mathrm{0.20}$Ge$_\mathrm{0.80}$&	GaAs&	InP&	Cubic Si\\ \hline
$0.7\times10^{-10}cm^3/s$-$11\times10^{-10}cm^3/s$&	
$3.5\times10^{-10}cm^3/s$&	$1.2\times10^{-10}cm^3/s$&	$4.73\times10^{-15}cm^3/s$\\
\end{tabular}
\end{ruledtabular}
\end{table}

\section{Comparison with previous generation Hex-Ge nanowire shells}\label{section9}
The evolution of the photoluminescence spectra for different growth recipes will be published on \url{https://researchdata.4tu.nl/} as open data sets. Briefly, the spectra of Hex-Ge started to be considerably broadened when they were grown by using a WZ-GaP core nanowire as shown in Supplementary Fig.~\ref{figs6}a In addition, we were not able to observe a clear spectrum at room temperature. The photoluminescence spectra improved by using a WZ-GaAs core to grow the Hex-Ge shell at a temperature of 600\,\textdegree C and 650\,\textdegree C, as shown in Supplementary Fig.~\ref{figs6}b, c respectively. The final improvement of the photoluminescence quality was realized by using longer WZ-GaAs core NWs, which yield the spectra in Fig.~\ref{fig3}a, ~\ref{fig3}b.

\end{document}